%% file: paper.tex
\documentclass[aps,prl,twocolumn,secnumarabic,notitlepage,superscriptaddress,nofootinbib,tightenlines,floatfix]{revtex4-1}

\usepackage[pdftex]{color}
\usepackage{graphicx,capt-of}
\usepackage{amsmath,amssymb,amsfonts,mathtools}
\usepackage[colorlinks=true,linkcolor=black,filecolor=black,urlcolor=blue,citecolor=blue,pdftex,plainpages=false]{hyperref}

\newcommand{\Gnorm}{G^\mathrm{norm}}
\newcommand{\tauf}{\tau_\mathrm{F}}
\DeclareMathOperator*{\SumInt}{%
    \mathchoice%
    {\ooalign{$\displaystyle\sum$\cr\hidewidth$\displaystyle\int$\hidewidth\cr}}
    {\ooalign{\raisebox{.14\height}{\scalebox{.7}{$\textstyle\sum$}}\cr\hidewidth$\textstyle\int$\hidewidth\cr}}
    {\ooalign{\raisebox{.2\height}{\scalebox{.6}{$\scriptstyle\sum$}}\cr$\scriptstyle\int$\cr}}
    {\ooalign{\raisebox{.2\height}{\scalebox{.6}{$\scriptstyle\sum$}}\cr$\scriptstyle\int$\cr}}
}
\begin{document}
\title{Heavy Quark Diffusion from 2+1 Flavor Lattice QCD with 320 MeV Pion Mass}
\author{Luis Altenkort}
\email{altenkort@physik.uni-bielefeld.de}
\affiliation{Fakult\"at f\"ur Physik, Universit\"at Bielefeld, D-33615 Bielefeld, Germany}
\author{Olaf Kaczmarek}
\affiliation{Fakult\"at f\"ur Physik, Universit\"at Bielefeld, D-33615 Bielefeld, Germany}
\author{Rasmus Larsen}
\affiliation{Department of Mathematics and Physics, University of Stavanger,
Stavanger, Norway}
\author{Swagato Mukherjee}
\affiliation{Physics Department, Brookhaven National Laboratory, Upton, New York 11973, USA}
\author{Peter Petreczky}
\affiliation{Physics Department, Brookhaven National Laboratory, Upton, New York 11973, USA}
\author{Hai-Tao Shu}
\email{hai-tao.shu@ur.de}
\affiliation{Institut f\"ur Theoretische Physik, Universit\"at Regensburg, D-93040 Regensburg, Germany}
\author{Simon Stendebach}
\affiliation{Insitut f\"ur Kernphysik, Technische Universit\"at Darmstadt, Schlossgartenstra{\ss}e 2, D-64289 Darmstadt, Germany}

\collaboration{HotQCD Collaboration}
\begin{abstract}
We present the first calculations of the heavy flavor diffusion coefficient using lattice QCD with light dynamical quarks corresponding to a pion mass of around 320 MeV. For temperatures $195\ \mathrm{MeV}<T<352\ \mathrm{MeV}$, the heavy quark spatial diffusion coefficient is found to be significantly smaller than previous quenched lattice QCD and recent phenomenological estimates. The result implies very fast hydrodynamization of heavy quarks in the quark-gluon plasma created during ultrarelativistic heavy-ion collision experiments. 
\end{abstract}
\date{\today}
\maketitle
\emph{Introduction.--} 
Heavy charm and bottom quarks are produced only during the earliest stages of ultrarelativistic heavy-ion collisions. They participate in the entire evolution of the quark-gluon plasma (QGP), and emerge as open heavy-flavor hadrons or quarkonia. These heavy-flavor hadrons provide valuable insights into the QGP. Experiments at the relativistic heavy ion collider (RHIC) and large hadron collider (LHC) show that the yields of heavy flavor hadrons at low transverse momentum ($p_T$) are  asymmetric along the azimuthal angle, meaning that the elliptic flow parameter ($v_2$) is large~\cite{Rapp:2018qla,Dong:2019byy,He:2022ywp}. These observations indicate that heavy quarks participate in the hydrodynamic expansion of the QGP. Understanding the origin of the hydrodynamic behavior of heavy quarks is key to understanding the near-perfect fluidity of the QGP. 

The motion of a heavy quark with mass $M\gg T$, immersed in a QGP at temperature $T$, can be effectively described by Langevin dynamics~\cite{Moore:2004tg}. The heavy quark momentum diffusion coefficient, $\kappa$, quantifies the momentum transfer to the heavy quark from the QGP background through random momentum kicks which are uncorrelated in time. Specifically, $3\kappa$ is the mean squared momentum transfer per unit time. For sufficiently low $p_T$, hydrodynamization of heavy quarks in the QGP can be characterized by the diffusion constant in space $D_s=2T^2/\kappa$~\cite{Moore:2004tg}, where $6D_s$ is the mean squared distance traversed per unit time by the heavy quark inside the QGP. 
The available perturbative QCD results for  $\kappa$~\cite{CaronHuot:2007gq} provide reliable estimates only for asymptotically large $T$. For practical phenomenological studies of heavy-ion collisions lattice QCD results for $\kappa$ are needed. Until now, lattice QCD based determinations of $\kappa$ have been limited only to pure gauge theory, therefore neglecting dynamical fermions entirely. Here, we report the first continuum-extrapolated lattice QCD calculations of $\kappa$ in 2+1-flavor QCD with a physical strange quark mass and degenerate up and down quark masses corresponding to a pion mass $m_\pi\simeq 320$~MeV.

\emph{Theoretical framework.--}
The momentum diffusion coefficient $\kappa$ can be obtained from the zero-frequency limit of the spectral function corresponding to the conserved current-current correlation function of heavy quarks. Relying on $M\gg T$ and $M \gg \Lambda_\mathrm{QCD}$, the QCD Lagrangian can be expanded in powers of $1/M$.  By integrating out the heavy quark fields one arrives at the heavy quark effective theory. In this effective theory the heavy-quark current-current correlator at leading order
in $1/M$ becomes equivalent to the correlation function of the colorelectric field~\cite{CasalderreySolana:2006rq,CaronHuot:2009uh}:
\begin{equation}
    \label{eq:gelat}
    G_E = -\sum_{i=1}^{3} 
    \frac{\left\langle {\rm Re Tr}\left[U(1/T,\tau)E_i(\tau,\mathbf{0})U(\tau,0)E_i(0,\mathbf{0})\right]\right\rangle}{3\left\langle {\rm Re Tr} U(1/T,0)\right\rangle} \,. 
\end{equation}
Here, $U(\tau_1,\tau_2)$ is the temporal Wilson line between Euclidean time $\tau_1$ and $\tau_2$, and $E_i(\mathbf{x},\tau) = U_i(\mathbf{x},\tau) U_4(\mathbf{x}+\hat{i},\tau) - U_4(\mathbf{x},\tau)U_i(\mathbf{x}+\hat{4})$ is the discretized chromo-electric field~\cite{CaronHuot:2009uh}.
$G_E$ receives only a finite renormalization at nonzero lattice spacing~\cite{Christensen:2016wdo}. In the continuum limit
the corresponding spectral function, $\rho_E(\omega,T)$, can be obtained by inverting~\cite{CaronHuot:2009uh}:
\begin{equation}
    G_E(\tau,T)=\int_0^{\infty}\frac{\mathrm{d} \omega}{\pi}\ \rho_E(\omega,T) \frac{\cosh[\omega\tau-\omega/(2T))]}{\sinh[\omega/(2T)]} \,, 
\label{spectral}
\end{equation}
where
\begin{equation}
    \kappa(T)=2 T \lim_{\omega \to 0} \left[ \rho_E(\omega,T)/\omega \right] \,,
\end{equation}
up to corrections proportional to $T/M$. 

The leading order (LO)~\cite{CaronHuot:2009uh} and
next-to-leading order (NLO)~\cite{CaronHuot:2009uh} perturbative QCD estimates predict $\rho_E(\omega,T)\propto\omega^3$, which should be valid for sufficiently large $T$ and/or $\omega$. Therefore, $G_E(\tau,T)$ is expected to receive significant contributions also from the high-frequency regions.

\emph{Lattice QCD calculations.--} 
We performed calculations in 2+1 flavor QCD with a physical strange quark mass, $m_s$, and degenerate up, down quark masses $m_l=m_s/5$ using the highly improved staggered Quark (HISQ) action~\cite{Follana:2006rc} and tree-level improved Lüscher-Weisz gauge action~\cite{Luscher:1984xn,Luscher:1985zq}. In the continuum limit our choice of $m_l$ corresponds to $m_\pi\simeq 320$ MeV. The lattice spacing $a$ and the quark masses are fixed as in Refs.~\cite{HotQCD:2014kol,Bazavov:2017dsy}. We carried out calculations on $96^3 \times N_{\tau}$ lattices with $1/a=7.036\,\mathrm{GeV}$ and $N_{\tau}=20,~24,~28,~32$ and $36$, that correspond to temperatures $T=352,~293,~251,~220$ and $195$ MeV, respectively. To control discretization effects we also performed calculations on $64^3 \times N_{\tau}$ lattices ($N_{\tau}=20,~22$ and $24$) at different lattice spacings, chosen such that the above temperature values are reproduced. Further details on the lattice setup are given in the Supplemental Material~\cite{supplemental}.

\begin{figure}[tbh]
    \centering
    \includegraphics{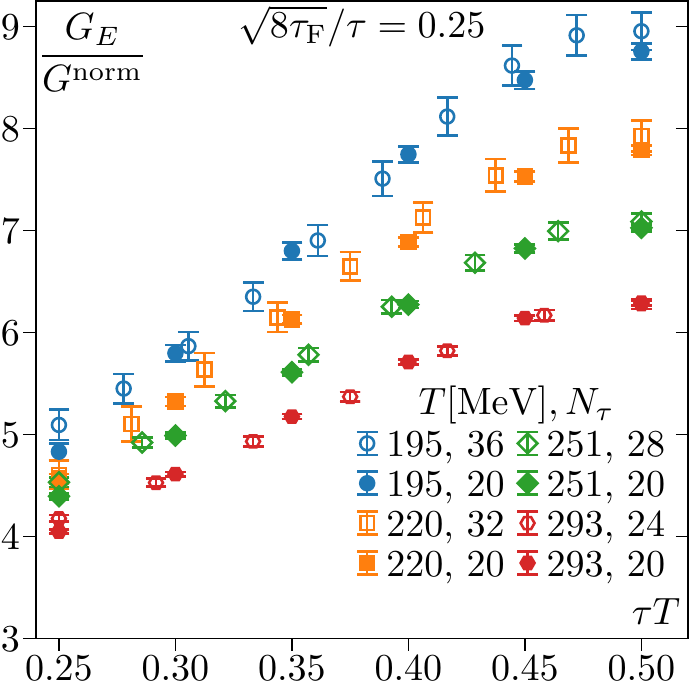}\\[4mm]
    \includegraphics{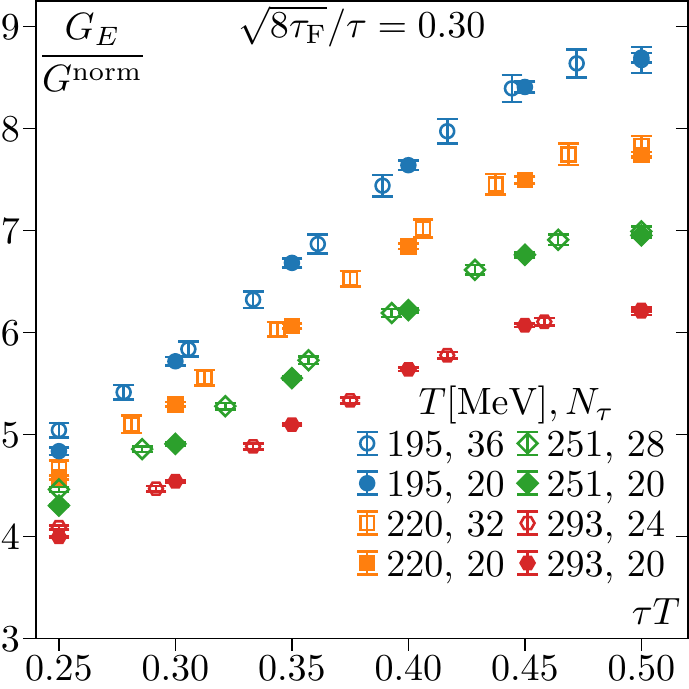}
    \caption{The chromoelectric correlator normalized by its weak-coupling structure at tree level ($\Gnorm$) as a function of $\tau$
    calculated on the $96^3 \times N_{\tau}$ lattices (open symbols) and $64^3 \times 20$ lattices (filled symbols) at two different flow times in units of $\tau$.}
    \label{fig:rawdata}
\end{figure}

Naive measurements of $G_E(\tau,T)$ are highly susceptible to high-frequency fluctuations in the gauge fields and exhibit a poor signal-to-noise ratio.
In quenched QCD, the multilevel algorithm~\cite{Luscher:2001up} has been applied to overcome this problem. However, this algorithm is not applicable for QCD with dynamical fermions. To overcome the noise problem for our calculations with dynamical fermions we use the Symanzik-improved~\cite{Ramos:2015baa} gradient flow~\cite{Luscher:2010iy}. In quenched QCD it was demonstrated~\cite{Altenkort:2020fgs, Brambilla:2022xbd} that this approach is as effective as the multi-level algorithm for noise reduction, while also renormalizing $G_E$ nonperturbatively. By evolving the gauge fields in the fictitious flow time, $\tauf$, as dictated by the force given by the gradient of the gauge action, the gradient flow smears the gauge fields over the radius $\sqrt{8 \tauf}$. Renormalization artifacts of the electric field operators $E_i$ due to finite lattice spacing $a$ are highly suppressed for $\sqrt{8 \tauf}>a$. However, the flow radius should always be smaller than the relevant physical scales, implying the constraint $\sqrt{8 \tauf}<\tau<1/(2T)$.

For $G_E$ it was found that the more strict criterion $\sqrt{8 \tauf}/\tau <1/3$ should be respected~\cite{flowlimits, Altenkort:2020fgs,Brambilla:2022xbd}. 

\emph{Results.--}
Since $\rho_E\propto\omega^3$ for large $\omega$, $G_E$ is a steeply falling function of $\tau$. Therefore, it is convenient to normalize it by the leading-order perturbative result~\cite{CaronHuot:2009uh}, with the Casimir factor, $C_F=4/3$, and the coupling constant, $g$, scaled out, that is, with $\Gnorm\equiv G_E^\mathrm{LO}/(g^2C_F)$.

At small $\tau$ the lattice results will suffer from significant discretization effects. 
Furthermore, the distortions of the correlation functions due to gradient flow are the largest at
small $\tau$. The cutoff effects as well as the distortions due to gradient flow
are also present in the free field theory. We can use the free theory result to estimate
and partly correct for these effects.
To reduce lattice cutoff effects as well as distortions due to gradient flow we perform tree-level improvement, meaning that we 
multiply the chromoelectric correlator by the ratio of the free correlator obtained in the continuum 
and the one calculated on the lattice (in perturbation theory) with the given $N_{\tau}$ at nonzero flow time:
$G_E(\tau,T) \rightarrow G_E(\tau,T) \times [\Gnorm(\tau T)/\Gnorm_{\tauf}(\tau T,N_{\tau})]$. The details of calculating $\Gnorm_{\tauf}(\tau T,N_{\tau})$ can be found in Supplementary material \cite{supplemental}.

The lattice chromo-electric correlators after tree-level improvement and normalized by $\Gnorm$ for the $96^3 \times N_{\tau}$ lattices are shown in Fig.~\ref{fig:rawdata}. We show the results for two different amounts of gradient flow, adjusted for each separation $\tau$ by fixing $\sqrt{8\tauf}/\tau$. In addition, we show the results from the coarsest lattices ($64^3 \time 20$) as open symbols. We see that gradient flow is effective in reducing UV noise
even for the largest lattice with $N_{\tau}=36$. After tree-level improvement, the difference of $G_E/\Gnorm$ obtained on the finest and the coarsest lattice is generally smaller than the statistical errors of the data obtained on the finest lattice. The flow time dependence is also quite small for $\tau T>0.25$ if the ratio of the flow radius $\sqrt{8 \tauf}$ to $\tau$ is between $0.25$ and $0.3$. For $\tau T<0.25$ the amount of flow necessary to suppress discretization artifacts already comes close to the relevant physical scale of $\tau/3$, leading to large distortions. For this reason the corresponding data points need to be omitted from the analysis. 

Naively one expects that at high (but not extremely high) temperatures $G_E/\Gnorm$ should not be
different from unity since $G_E^\mathrm{LO}\approx g^2 C_F$ is a good approximation for $G_E$ and
$C_F g^2 \simeq 1$. 
An interesting feature of the results shown in Fig. \ref{fig:rawdata} is
that the ratio $G_E/\Gnorm$ has a much larger deviation from one than in the quenched case. 
In quenched QCD, $G_E/\Gnorm$ reaches a value of about $4$ at most \cite{Brambilla:2020siz}. This is due to the fact that the $\tau$ values in physical units ($\mathrm{fm}$) accessible in full QCD are larger and, as we will see later, the value of $\kappa$ in temperature units also turn out to be larger. As in quenched QCD, deviations from unity of $G_E/\Gnorm$ are the largest at the lowest temperature, and become smaller as the temperature increases.

\begin{figure}
    \centering
    \includegraphics{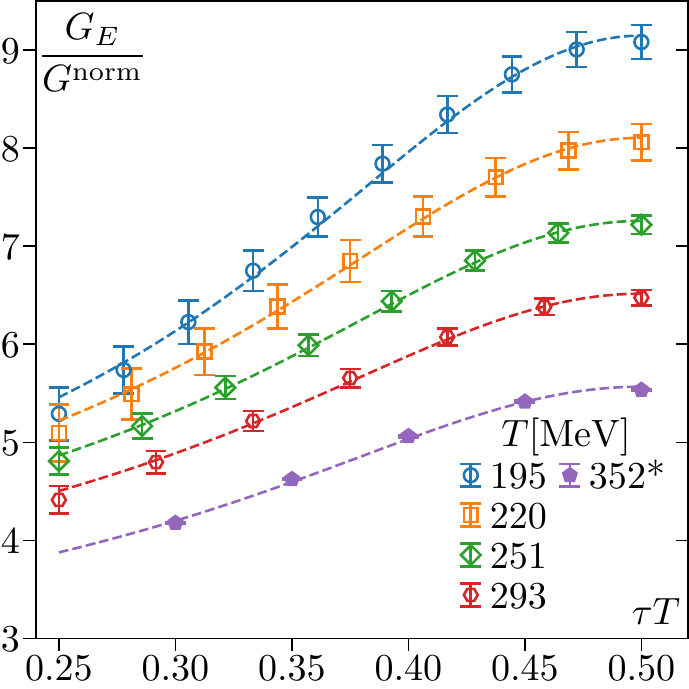}
    \caption{The continuum and zero flow time extrapolated results for the chromo-electric correlator at different
    temperatures as a function of $\tau T$. Also shown is the result for the highest temperature at nonzero lattice spacing corresponding to $N_\tau=20$ and 
    flow time $\sqrt{8\tau_\mathrm{F}}/\tau=0.3$. The dashed lines indicate fitted model correlators for the ``smax'' model using the NLO $\rho_\mathrm{UV}$.}
    \label{fig:contres}
\end{figure}

Next, we perform the continuum and flow-time-to-zero extrapolation of the chromoelectric correlator. First, we interpolate the correlators obtained on $64^3 \times N_{\tau}$ lattices in $\tau$ for various values of $\tauf/\tau^2$. From these interpolations we determine the correlator on coarser lattices for values of $\tau$ that are available for the finest $96^3 \times N_{\tau}$ lattices and then perform continuum extrapolations for each $\tau T$ and $\tauf/\tau^2$. As is apparent from Fig.~\ref{fig:rawdata}, the cutoff effects are small except for small values of $\tau T$. We perform the continuum extrapolation assuming that discretization errors go like $(a T)^2\sim 1/N_{\tau}^2$, which turns out to be capable of describing our data well, see Supplemental Material~\cite{supplemental}.

Finally, we perform the flow-time-to-zero extrapolation of the chromoelectric correlators. In the region $a \ll \sqrt{8 \tauf} \ll \tau$ we expect a  linear $\tauf$ dependence as suggested by NLO perturbation theory \cite{Eller:2021qpp}. And indeed, for $0.25< \sqrt{8 \tauf}/\tau < 0.3$ a linear dependence seems to describe the data. Therefore, we use a linear extrapolation
in $\tau_F$ in this region to obtain the zero flow time limit, see Supplemental Material \cite{supplemental}.
The continuum and zero flow time extrapolated results for the chromoelectric correlators are shown in Fig. \ref{fig:contres}. 
The extrapolations do not change the qualitative features of the correlation function but lead to a significant increase of the statistical errors.

With the continuum and flow-time-extrapolated data for the chromoelectric correlator we are in the position to estimate
the heavy quark diffusion coefficient $\kappa$. To do so we need a parametrization of the spectral function that enters Eq. (\ref{spectral}). Any parametrization of the spectral function should take into account its known behavior at
small and large $\omega$. For small $\omega$ the spectral function is solely determined by the heavy quark diffusion coefficient and has the form \cite{CaronHuot:2009uh}:
$\rho_E(\omega,T) \simeq \rho_\mathrm{IR}(\omega,T)=\kappa \omega/(2T)$
while at sufficiently large 
frequency the $\omega$ dependence of the spectral function should be described by perturbation theory due to asymptotic freedom in 
QCD. Moreover, thermal corrections to the spectral function are very small
for $\omega \gg T$. Therefore, we assume that at large energies the spectral
function is given by the LO or NLO perturbative $T=0$ result up to a constant:
$\rho_E(\omega\gg T)=\rho_\mathrm{UV}(\omega)=K \rho_\mathrm{LO,NLO}(\omega)$.
The factor $K$ accounts for the fact that the perturbative calculations may not be quantitatively reliable due to missing contributions from higher orders.

Perturbative calculations at NLO \cite{CaronHuot:2009uh,Burnier:2010rp},
classical simulations in effective three-dimensional theory \cite{Laine:2009dd}
and strong coupling calculations \cite{Gubser:2006nz} show that
the spectral function is a smooth monotonically rising function of $\omega$.
Based on this, as well as the above considerations, we use the following two forms of the spectral function in our analysis that also have been used already in quenched
QCD \cite{Francis:2015daa,Brambilla:2020siz}:
$\rho_{\mathrm{max}}=\max \Big{(}\rho_\mathrm{IR}(\omega,T),\rho_\mathrm{UV}(\omega)\Big{)}$
and $\rho_{\mathrm{smax}}=\sqrt{\rho_\mathrm{IR}^2(\omega,T)+\rho_\mathrm{UV}^2(\omega)}$,
which we refer to as the maximum ($\mathrm{max}$) and the smooth maximum ($\mathrm{smax}$) \emph{Ans\"atze}, respectively.
The latter is consistent with the perturbative NLO calculation \cite{Burnier:2010rp}
and OPE considerations \cite{Caron-Huot:2009ypo} when it comes to the leading thermal correction at $\omega \gg T$.
We also consider a third \emph{Ansatz} for the spectral function, that is given by $\rho_\mathrm{IR}(\omega,T)$ up to $\omega=\omega_{\mathrm{IR}}$, and by $\rho_\mathrm{UV}$
for $\omega>\omega_\mathrm{UV}$, and for $\omega_\mathrm{IR}<\omega<\omega_\mathrm{UV}$ we interpolate with a power-law form $\rho(\omega,T)=c\ \omega^p$. The parameters $c$ and $p$ 
are chosen such that the spectral function is continuous at $\omega=\omega_\mathrm{IR}$
and $\omega_\mathrm{UV}$. This form of the spectral function has been used in Ref.~\cite{Brambilla:2020siz}. Based on theoretical results we choose $\omega_\mathrm{IR}=T$ and $\omega_\mathrm{UV}=2 \pi T$, see Supplemental Material \cite{supplemental}.

Using the above three \emph{Ans\"atze} for the spectral functions and the spectral
representation of the chromoelectric correlator we fitted the continuum-
and flow-time-extrapolated results treating $\kappa$ and $K$ as fit parameters
and thus estimated the heavy quark diffusion coefficient. 
It turns out that the maximum \emph{Ansatz} gives the largest value of $\kappa$, while the power-law
form gives the smallest value. Using the LO or NLO form of $\rho_\mathrm{UV}$ does
not lead to significant change in the value of $\kappa$, meaning that the 
estimated values of $\kappa$ are not too sensitive to the modeling of the high
energy part of the spectral function.

Each model is fitted onto the same 1000 bootstrap samples of the double-extrapolated correlator data. We collect all results from all models in a single ``distribution'' for the fit parameter $\kappa/T^3$. We determine a confidence interval by considering the median of this distribution, and then adding or subtracting the 34th percentiles on each side, which gives the lower and upper bounds of the interval. For better readability we quote the central value of the interval with the distance to the bounds as the uncertainty. We obtain:
$\kappa(T=195\,\mathrm{MeV})= 11.0(2.5) T^3,~\kappa(T=220\,\mathrm{MeV})= 8.4(2.4) T^3,~\kappa(T=251\,\mathrm{MeV})= 6.9(2.2) T^3$, and $\kappa(T=293\,\mathrm{MeV})= 5.8(2.0) T^3$.
\begin{figure}
    \includegraphics[width=8cm]{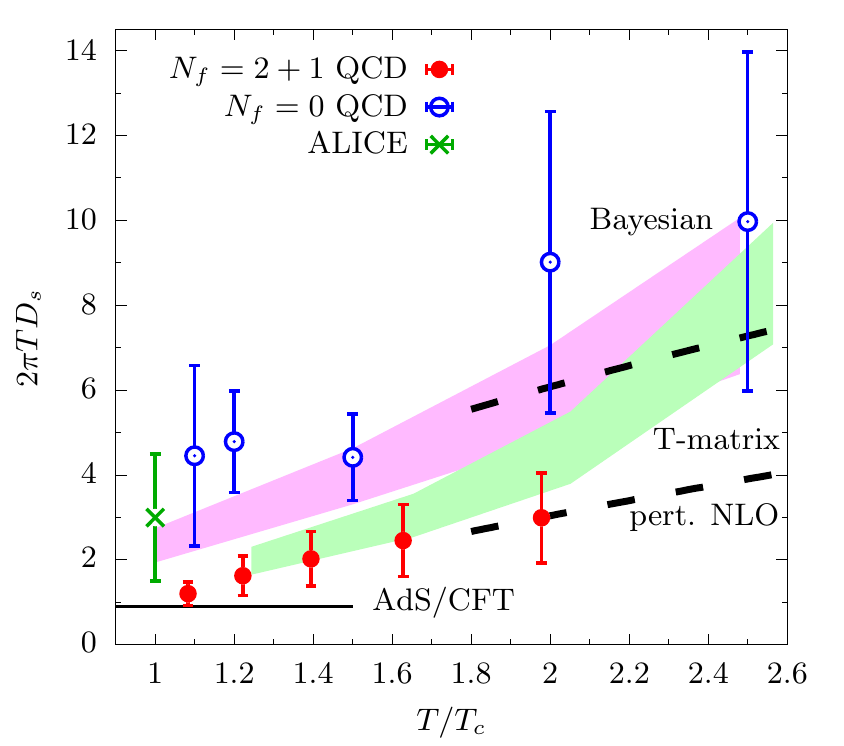}
    \caption{The spatial heavy quark diffusion coefficient in units of $2 \pi T$
    from our lattice calculations compared to the AdS/CFT estimate \cite{CasalderreySolana:2006rq}, NLO perturbative calculation \cite{CaronHuot:2007gq} and the 
quenched lattice QCD calculations \cite{Altenkort:2020fgs,Brambilla:2020siz,Banerjee:2022gen}.
For the quenched lattice data we show the result of Ref.~\cite{Brambilla:2020siz} for $T=1.1T_c$,
the result of Ref.~\cite{Altenkort:2020fgs} for $T=1.5T_c$, and the results of Ref.~\cite{Banerjee:2022gen} for the remaining temperatures. 
For the NLO calculations we used two values of the renormalization
scale, $\mu=2 \pi T$ (lower dashed line) and $\mu=4 \pi T$ (upper dashed
line). Also shown are the phenomenological estimates \cite{Liu:2016ysz,Liu:2017qah,Xu:2017obm,ALICE:2021rxa}, see main text.}
\label{fig:twopiTD}
\end{figure}

As already noted above, the shape of the correlation function for $\tau T>0.25$ does not seem to be significantly changed by the continuum and flow-time-to-zero extrapolations.
Therefore, we also performed the above analysis using the nonzero lattice spacing data at $1/a=7.036\,\mathrm{GeV}$ and nonzero relative flow times $\sqrt{8 \tauf}/\tau=0.3$. We find that the estimated values of 
$\kappa$ agree 
with the ones obtained from the continuum- and flow-time-extrapolated data 
within errors. This is due to the fact that the systematic uncertainties
associated with modeling of the spectral function are much larger than the effect of the continuum and zero-flow-time extrapolation. For this reason we also estimate the heavy quark diffusion coefficient at nonzero lattice spacing and flow time for the highest
temperature resulting in $\kappa(T=356~{\rm MeV})/T^3=4.8(1.7)$.

\emph{Conclusion.--} 
We carried out first lattice QCD calculations of the heavy quark diffusion constant in 2+1 flavor QCD at leading order in the inverse heavy quark mass and in the phenomenologically relevant region of 195 MeV $<T<$ 352 MeV. Our results for $D_s$ as function of $T/T_c$ are summarized in Fig.~\ref{fig:twopiTD}. Here we use $T_c=180$ MeV because the calculations
are performed at $m_{\pi}\simeq 320$ MeV, see Supplemental Material~\cite{supplemental}. Our results are smaller than the quenched lattice QCD results~\cite{Brambilla:2020siz,Banerjee:2022gen}. At the lowest temperature our result agrees, within errors, with the strong coupling expectations from AdS/CFT~\cite{CasalderreySolana:2006rq,Andreev:2017bvr}. At the highest temperature our result $D_s$ is compatible with the NLO perturbative prediction~\cite{CaronHuot:2007gq} within
the uncertainties.  In comparisons to some phenomenological determinations, namely, the Bayesian analysis~\cite{Xu:2017obm} of heavy-ion collision data, and the ALICE Collaboration's model fits to their data~\cite{ALICE:2021rxa}, the lattice QCD results for $D_s$ are systematically smaller. On the other hand the the $T$-matrix approach on $D_s$ \cite{Liu:2016ysz,Liu:2017qah} seems to agree with the lattice results.

The present study can be extended in two different ways. Based on the previous lattice QCD 
studies of QCD equation of state~\cite{Bazavov:2017dsy}, quark number susceptibilities~\cite{Bazavov:2010bx,Bazavov:2010sb}, and static quark free energies~\cite{Bazavov:2018wmo} with light quark mass $m_l=m_s/5$ compared to those with nearly physical light quark mass $m_l\simeq m_s/20$, we expect the effect larger than physical quark mass on $D_s$ to be small in the temperature range considered in this study. However, such effects might be significant for temperatures closer to the QCD crossover temperature. Thus, we plan to extend the present calculations to smaller temperatures with physical values of the light quark masses. The heavy quark mass suppressed effects to the heavy quark diffusion coefficients are expected to be relatively small based on the calculations performed in quenched QCD \cite{Banerjee:2022uge,Brambilla:2022xbd}. We plan to estimate these corrections also in 2+1 flavor QCD.

All data from our calculations, presented in the figures of this paper, can be found in \cite{datapublication}.

The computations in this work were performed using \texttt{SIMULATeQCD}~\cite{Mazur:2021zgi, Bollweg:2021cvl}. Parts of the computations in this work were performed on the GPU cluster at Bielefeld University. We thank the Bielefeld HPC.NRW team for their support. 

%
%
This material is based upon work supported by The U.S. Department of Energy, Office of Science, Office of Nuclear Physics through Contract No.~DE-SC0012704, and within the frameworks of Scientific Discovery through Advanced Computing (SciDAC) award \textit{Fundamental Nuclear Physics at the Exascale and Beyond} and the Topical Collaboration in Nuclear Theory \textit{Heavy-Flavor Theory (HEFTY) for QCD Matter}.
R.L. acknowledge funding by the Research Council of Norway under the FRIPRO Young Research Talent Grant No. 286883. L.A., O.K. and S.S. acknowledge support by the Deutsche For\-schungs\-ge\-mein\-schaft
(DFG, German Research Foundation) through the CRC-TR 211 ``Strong-interaction matter under extreme conditions"– Project No. 315477589 – TRR 211. 

This research used awards of computer time provided by the National Energy Research Scientific Computing Center (NERSC), a U.S. Department of Energy Office of Science User Facility located at Lawrence Berkeley National Laboratory, operated under Contract No. DE-AC02- 05CH11231, and the PRACE awards on JUWELS at GCS@FZJ, Germany and Marconi100 at CINECA, Italy. Computations for this work were carried out in part on facilities of the USQCD Collaboration, which are funded by the Office of Science of the U.S. Department of Energy.

We thank the Institute for Nuclear Theory at the University of Washington for its kind hospitality and stimulating research environment during the completion of this work. INT is supported in part by the U.S. Department of Energy Grant No. DE-FG02- 00ER41132.

We thank Guy D. Moore for valuable discussions.

\newpage
\input{supplementary.tex}
\end{document}

%% file: supplementary.tex
\def\figureautorefname{Fig.}
\def\tableautorefname{Tab.}
\def\equationautorefname{Eq.}

\begin{widetext}
\vspace{2em}
\begin{center}
    {\Large \bf Supplemental materials}
\end{center}
\vspace{2em}

In these supplemental materials we discuss important technical details on 
the calculation of the chromo-electric correlator and the heavy quark diffusion coefficient
presented in the letter.
In section \ref{supp-mat:setup} we discuss additional details about the lattice setup. In section \ref{supp-mat:LO-lattice} we discuss the calculation
of the free chromo-electric correlator (leading-order result) on the lattice
with gradient flow. In section \ref{supp-mat:extrap}, we discuss the continuum
and flow time extrapolations of the lattice results. Finally, in section
\ref{supp-mat:SPF}, we discuss the reconstruction of the spectral function and the
determination of the heavy quark momentum diffusion coefficient.

\section{Lattice setup}
\label{supp-mat:setup}

\begin{table}[b]
\begin{minipage}[t]{0.45\textwidth}
    \centering
    \begin{tabular}{|c|ccccc|}
    \hline
    $N_{\tau}$  & 36   & 32   & 28   & 24   & 20  \\
    \hline
    $T$ [MeV]   & 195  & 220  & 251  & 293  & 352 \\
    \hline
    \# conf.    & 2256 & 912 & 1680 & 688 & 2488 \\
    \hline
    \end{tabular}
    \caption{Parameters of the lattice calculations on the finest lattices ($96^3\times N_\tau$) with bare parameters $\beta=8.249$, $a m_s=0.01011$, $m_l=a m_s/5=0.002022$.}
    \label{tab:paramf}
\end{minipage}\hfill
\begin{minipage}[t]{0.45\textwidth}
        \centering
\begin{tabular}{|c|ccccc|}
\hline
$T$ [MeV] & $\beta$ & $a m_s$ & $a m_l$ & $N_{\tau}$ & \# conf.\\
\hline
195 & 7.570	& 0.01973 &	0.003946 & 20 & 5899 \\
    & 7.777	& 0.01601 &	0.003202 & 24 & 3435 \\
\hline
220 & 7.704	& 0.01723 &	0.003446 & 20 & 7923 \\
    & 7.913	& 0.01400 &	0.002800 & 24 & 2715 \\
\hline 
251 & 7.857	& 0.01479 &	0.002958 & 20 & 6786 \\
    & 8.068	& 0.01204 &	0.002408 & 24 & 5325 \\
\hline
293 & 8.036	& 0.01241 &	0.002482 & 20 & 6534 \\
    & 8.147	& 0.01115 &	0.002230 & 22 & 9101 \\
    \hline
\end{tabular}
\caption{Parameters of the lattice calculations on the coarse lattices ($64^3\times N_\tau$), including the temperatures, the gauge couplings,
the bare quark masses in lattice units, the values of $N_{\tau}$, and the corresponding statistics (last column).}
\label{tab:paramc}
    \end{minipage}
\end{table}

As mentioned in the paper,
we performed lattice QCD calculations of the chromo-electric correlator 
using the HISQ action for quarks and the Symanzik-improved L\"uscher-Weisz action for gluons.
The calculations have been performed
at the physical
strange quark mass and light quark masses $m_l=m_s/5$ in a fixed scale approach on $96^3 \times N_{\tau}$
lattices for the bare gauge coupling $\beta=10/g_0^2=8.249$, 
corresponding to a lattice spacing of $1/a=7.036$ GeV. The corresponding bare strange strange quark mass
for this value of $\beta$ in lattice units is $a m_s=0.01011$ \cite{HotQCD:2014kol}.
The temperature was varied by using $N_{\tau}=36,~32,~28,~25$ and $20$, corresponding
to temperatures $T=195,~220,~251,~293$ and $352$ MeV, respectively. The lattice spacing
and thus the temperature scale has been fixed using the $r_1$-scale determined in Ref.~\cite{Bazavov:2017dsy} with the value $r_1=0.3106\,\mathrm{fm}$ obtained in Ref.~\cite{MILC:2010hzw}.
The value of the strange quark mass was obtained from the parametrization of the line of
constant physics from Ref.~\cite{HotQCD:2014kol}. The parameters of these calculations, including
the statistics, are presented in \autoref{tab:paramf}. 

To control discretization effects and
perform continuum extrapolations for each temperature (except the largest one), we performed
calculations on two coarser lattices with lattice sizes $64^3 \times N_{\tau}$ and $N_{\tau}=20,~22/24$. Here, the temperature values were adjusted by varying the bare gauge 
coupling $\beta$. Again we used the $r_1$-scale to set the temperature
scale and lines of constant physics  to obtain the quark masses from
Ref. \cite{Bazavov:2017dsy} and \cite{HotQCD:2014kol}, respectively.
In \autoref{tab:paramc} we present
the lattice details of the corresponding  calculations including the temperatures, temporal lattice sizes $N_{\tau}$, quark masses and bare lattice gauge coupling, as well as the corresponding statistics.

The gauge configurations are generated using the RHMC algorithm. We save configurations every 10 trajectories and the acceptance rate is tuned to $\approx80\%$. 
To estimate the auto-correlation time we consider the expectation value of the Polyakov loop
evaluated on the flowed gauge configurations at the maximum flow time that enters the analysis, which is $\sqrt{8\tau_\mathrm{F}}T=0.15$.
The integrated autocorrelation times turn out to be around 10-30 saved configurations. In each stream we bin the data into bins with length of one integrated autocorrelation time and use the approximately independent bins for the bootstrap analysis.

To facilitate the comparison of our lattice results on $\kappa$ with available quenched
QCD results and some model calculations, it is worth to show them as a function of temperature $T$ in units of a (pseudo)critical transition temperature $T_c$, chosen appropriately for each case. 
In the quenched case, $T_c$ is the critical temperature of the first-order deconfinement phase transition. 
In full QCD, this transition is only well defined at significantly larger light quark masses than what is used in this study. However, for the case of 2+1-flavor QCD with $m_l=m_s/5$ the light quark mass is sufficiently small to associate $T_c$ with the chiral crossover temperature.
Typically the chiral crossover temperature is defined as the peak
position in the disconnected
chiral susceptibility. For the HISQ action with 2+1-flavor QCD and $m_l=m_s/5$
the disconnected chiral susceptibility has been calculated in Ref.~\cite{Bazavov:2010sb} on $24^3 \times 6$ lattices.
Using the corresponding data from Ref.~\cite{Bazavov:2010sb} and performing fits
with Gaussian and polynomial forms in the vicinity of the peak position,
we obtain $T_c=180(2)\,\mathrm{MeV}$, which we use to position the 2+1-flavor data in \autoref{fig:twopiTD}. The quoted error is statistical. The lower limit
of the fit range was set to be $T=170$ MeV, while the upper limit of the fit range
was varied from $T=195$ MeV to $T=210$ MeV. The peak position does not depend
on the fit range or on the form of the fit function (polynomial or Gaussian).

\section{Leading order correlation functions at finite flow time in lattice perturbation theory}
\label{supp-mat:LO-lattice}
In the following, we will lay out the calculation of the leading order lattice perturbation theory correlation function $G_{\tauf}^\mathrm{norm}(\tau T,N_\tau)$ under flow that is used for tree level improvement \cite{2022-Stendebach}. Under gradient flow, the lattice gluon propagator takes the form
\begin{align}
\begin{split}
    \left(D^\mathrm{flow}\right)_{\mu\nu}^{AB}(p,\tauf) & = \left(e^{-\tauf K^{(f)}(p)}\right)_{\mu\sigma}\times\\
    &\phantom{iii}\left(e^{-\tauf K^{(f)}(p)}\right)_{\nu\rho}D_{\sigma\rho}^{AD}(p)
\end{split}
\end{align}
where $K^{(f)}$ is the flow kernel and $D$ is the unflowed lattice propagator which is the inverse of the action kernel $K^{(a)}$. The matrices depend on the discretization of gradient flow and action respectively. For Zeuthen flow and the L\"uscher-Weisz gauge action, the kernels can be found in Ref.\ \cite{Ramos:2015baa}. Using the flowed propagator, leading order perturbation theory expectation values of some gluonic operator $M$ under flow can be written as 
\begin{equation}
    \left\langle M\right\rangle_0 = \SumInt_pM_{\mu\nu}^{AB}(p)\left(D^\mathrm{flow}\right)_{\mu\nu}^{AB}(p,\tauf),
    \label{eq:pert_observable}
\end{equation}
where $M_{\mu\nu}^{AB}$ is an observable-dependent matrix, and $\langle \dots \rangle_0$ means
expectation value in the free field theory.
The symbol $\SumInt_p$ is to be understood as a shorthand notation for $T\sum_{p_0}\int d^3p/(2\pi)^3$. By taking the lattice expression of the color-electric field strength $G_E$ and expanding the link variables $U_\mu(x)=\exp (iag_0A_\mu(x))$ in the Lie algebra fields, we find the operator matrix
\begin{equation}
    M_{\mu\nu}^{AB}(p,\tau) = -\frac{g_0^2}{18}\cos\left(p_0\tau\right)\delta^{AB}m_{\mu\nu}(p).
\end{equation}
Here,
\begin{align}
\begin{split}
    m_{00}(p) & = \sum_i\hat{p}_i^2,\\
    m_{ii}(p) & = \hat{p}_0^2,\\
    m_{0i}(p) & = -\hat{p}_0\hat{p}_i = m_{i0}(p),
\end{split}
\end{align}
with the lattice momentum $\hat{p}_\mu = (2/a)\sin (p_\mu a/2)$. Using this, Eq.\ (\ref{eq:pert_observable}) is integrated numerically for each $\tau$ and $\tauf$ to obtain
the leading-order result 
on the lattice, $G_{E,\tauf}^\mathrm{LO}(\tau T,N_\tau)$. The accuracy of the integration is tested by making use of the fact that for Wilson action and Wilson flow, an analytical solution to the integral exists \cite{Altenkort:2020fgs}. 
Numerical integration for the Wilson-Wilson case agrees with the analytical result up to $10^{-8}$ which is much smaller than the statistical uncertainty, so the integration procedure is well suited. The normalized lattice correlator is then
defined as 
$G_{\tauf}^\mathrm{norm}(\tau T,N_\tau)=G_{E,\tauf}^\mathrm{LO}(\tau T,N_\tau)/(C_F g_0^2)$.

\section{Continuum and flow time extrapolations}
\label{supp-mat:extrap}

\begin{figure*}[b]
    \hfill%
    \includegraphics[page=21]{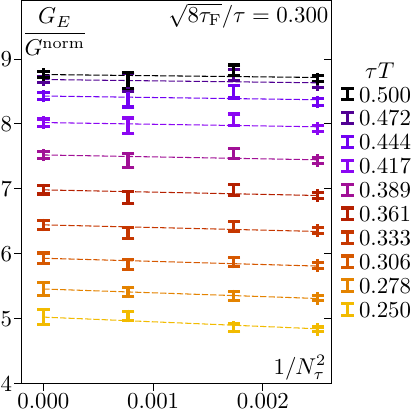}\hspace{2cm}
    \includegraphics[page=21]{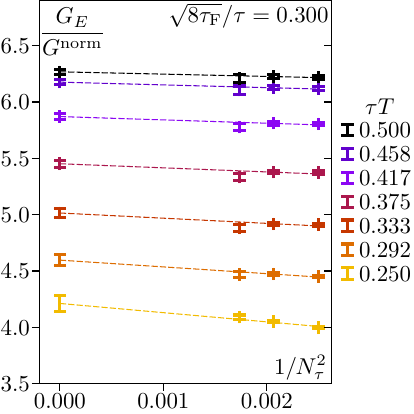}\hfill\null
    \caption{The lattice spacing dependence of the chromo-electric correlator at 
    $T=195\,\mathrm{MeV}$ (left) and $T=293\,\mathrm{MeV}$ (right) at $\sqrt{8\tauf}/\tau=0.3$. The dashed lines show the combined fit for the continuum extrapolations (see text) of the median values of the bootstrap samples.}
    \label{fig:cont_extra}
\end{figure*}

As in the main text in the following discussion we will normalize the chromo-electric correlator
by the free theory (LO) result with the Casimir factor $C_F=4/3$ and the coupling constant $g^2$
scaled out \cite{CaronHuot:2009uh}:
\begin{equation}
\Gnorm=\frac{G_E^\mathrm{LO}}{g^2 C_F}=\pi^2 T^4 \left[ \frac{\cos^2 \left(\pi \tau T \right)}{\sin^4\left(\pi \tau T \right) } + \frac{1}{3 \sin^2\left(\pi \tau T\right)} \right].
\end{equation}

To perform the continuum extrapolation we first use spline interpolations
of the $64^3 \times N_{\tau}$ ($N_{\tau}=20,~22/24$) lattice
data for $G_E$ in $\tau T$ at each temperature.
From the spline interpolations we determine $G_E$ at the 
values of $\tau T$ available on the finest ($96^3 \times N_{\tau}$)
lattices for the same temperature.
In this way we can study the lattice spacing dependence of $G_E/\Gnorm$
at each temperature in terms of $aT=1/N_{\tau}$. This is shown in
\autoref{fig:cont_extra} for $\sqrt{8 \tauf}/\tau = 0.3$ as an example. In general, we see that the lattice spacing ($N_{\tau}$) dependence of $G_E/\Gnorm$ is mild.

Unfortunately, even with gradient flow the statistical errors on the data are too large to perform naive individual continuum extrapolations for each distance $\tau T$, as these will suffer from overfitting of the noise. However, from our perturbative lattice calculations and from previous quenched results we expect the continuum to be approached from below, and we expect the slope of the extrapolation to vanish for very large $\tau T$, with a leading decay proportional to $1/(\tau T)^2$. Therefore, we constrain the slope of the extrapolation by performing a combined extrapolation of all $\tau T$ at fixed $\sqrt{8\tauf}/\tau$, with the Ansatz $f(N_\tau; \tau T)=\sum_{\tau T} \left( G^\mathrm{cont}_{\tau T} - (m/(\tau T))^2 N_\tau^{-2}\right)$. This is shown in \autoref{fig:cont_extra} for two temperatures at $\sqrt{8\tauf}/\tau=0.3$. For all temperatures and all relative flow times this fit ansatz yields a $\chi^2/\mathrm{d.o.f} \in [0.8, 1.9]$, meaning that the data is described well and that this constraint is not too restrictive given the statistical precision. To estimate the error on the continuum extrapolated results we use the bootstrap method.
\begin{figure*}[tb]
    \hfill%
    \includegraphics{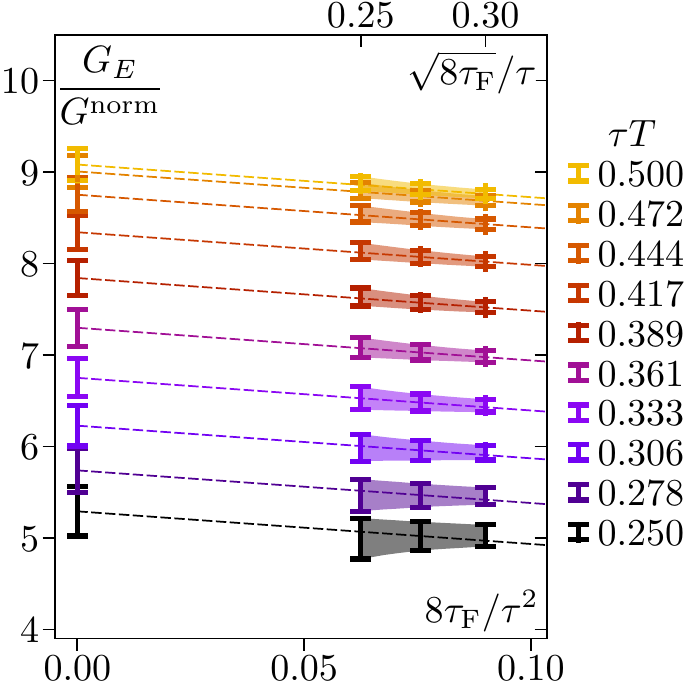}\hspace{2cm}
    \includegraphics{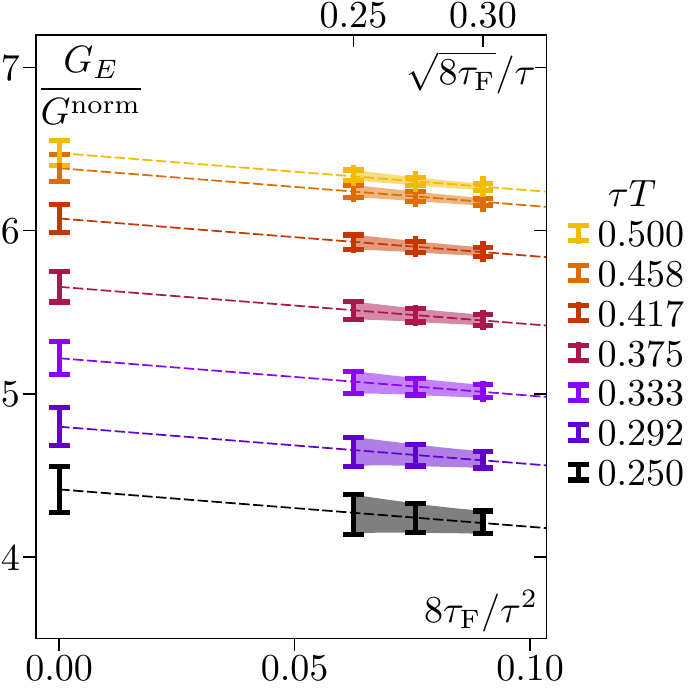}\hfill\null
    \caption{The flow time dependence of $G_E/\Gnorm$  for $T=195\,\mathrm{MeV}$ (left),
    and $T=293\,\mathrm{MeV}$ (right) shown as bands. The three explicitly shown data points correspond to the only values of $G_E/\Gnorm$ that enter in the combined flow time extrapolation (see text). The combined linear fit results are shown as dashed lines.
    }
    \label{fig:flow_extra}
\end{figure*}

Next we examine the flow time dependence of the continuum extrapolated
result for the chromo-electric correlator. 
The flow time dependence of the chromo-electric correlator is shown in \autoref{fig:flow_extra}.
As mentioned in the main text,
in the window $a \ll \sqrt{8 \tauf} \ll \tau/3$ we expect a  linear $\tauf$-dependence suggested by the NLO perturbative calculation \cite{Eller:2021qpp}. 
In the continuum extrapolated data we find that the upper bound for the flow time as suggested by perturbation theory is a bit too generous (meaning that nonlinear behavior starts already at slightly smaller flow times), which is why we restrict the upper limit to $ \sqrt{8 \tauf}/\tau < 0.3$. The lower bound is also chosen more conservatively, in order to be sure that discretization artifacts are well-suppressed and in order to obtain a good enough signal-to-ratio. For these reasons we perform the flow extrapolations in the range $0.25< \sqrt{8 \tauf}/\tau < 0.3$ for $\tau T\geq 0.25$. From our perturbative results, quenched studies and the nonzero lattice spacing data of this study, we expect the flow time limit to be approached from below, and we also expect the slope of the extrapolation to be roughly independent of $\tau T$. For these reasons, we again perform a combined extrapolation of all $\tau T$ with the Ansatz $f(\tauf; \tau T)=\sum_{\tau T}\left(G^{\tauf=0}_{\tau T} + m \, \tauf\right)$. Given our statistical precision in the continuum, this Ansatz can describe the data well for all temperatures.

We note that the results on $G_E$ evaluated for different flow times are strongly correlated. Therefore, for the zero flow time extrapolation
we restrict ourselves to use only three $\tauf$ values as indicated in \autoref{fig:flow_extra}.
Adding data on $G_E$ at more $\tauf$ values will not change the result of extrapolation.
We note that here we use uncorrelated fits because given our statistics it is impossible to reliably estimate the correlation matrix.
The errors on the extrapolated values are again estimated
using the bootstrap procedure.

We note that we also performed a second analysis for the continuum and flow-time-to-zero extrapolation, where the continuum
and flow time extrapolations are performed independently at each $\tau T$ value. 
In this analysis, to estimate the continuum limit we perform a constant fit of the $G_E/G^{norm}$ data for $\tau T>0.33$,
while for smaller $\tau T$ values we perform a fit with a linear Ansatz in $1/N_{\tau}^2$.
As explained above, we expect the flow time limit to be approached from below, which is why we constrain the slope of the extrapolation for each $\tau T$ to be smaller or equal zero.
The results for the continuum and zero flow time extrapolated correlator and $\kappa/T^3$ from this analysis agree with the above approach well within errors.

\section{Spectral reconstruction}
\begin{figure*}[p]
    \hfill
    \includegraphics{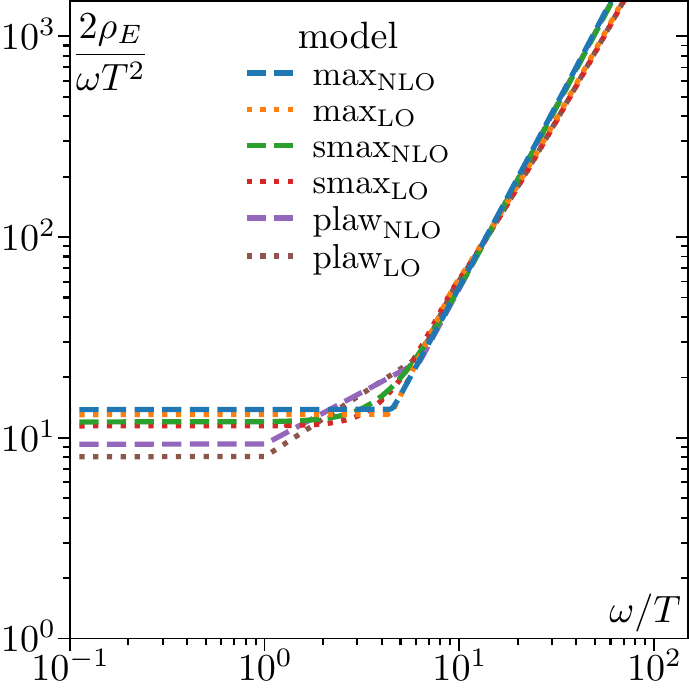}\hspace{2cm}
    \includegraphics{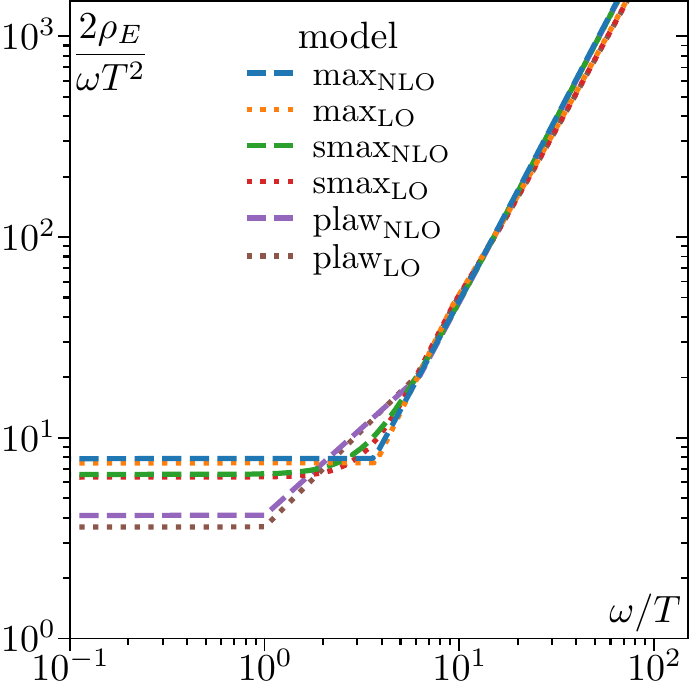}\hfill\null
    \caption{The spectral functions obtained for different fit forms for
    $T=195$ MeV (left) and $293$ MeV (right). Only the bootstrap median is shown; statistical errors are hidden for better visibility. The y-axis is scaled such that $\kappa/T^3$ can be read of at the y-intercept.}
    \label{fig:spf}
\end{figure*}
\begin{figure*}[p]
    \hfill
    \includegraphics{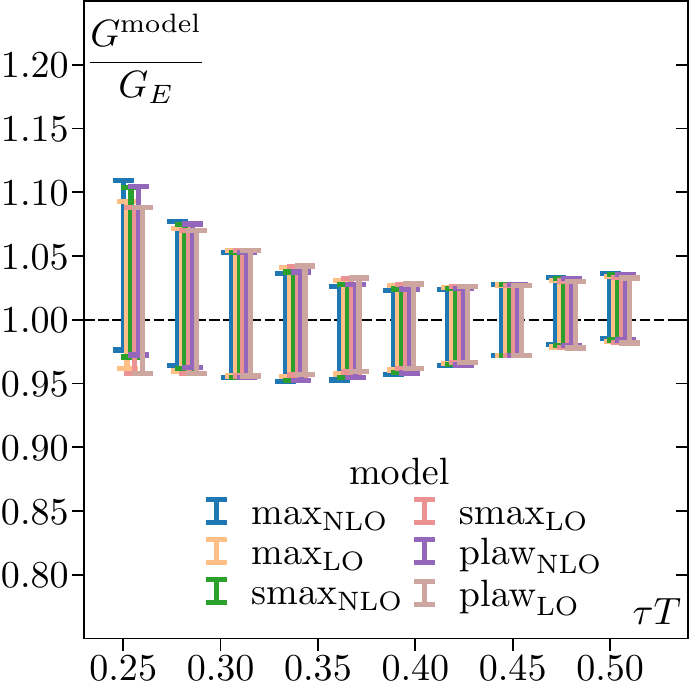}\hspace{2cm}%
    \includegraphics{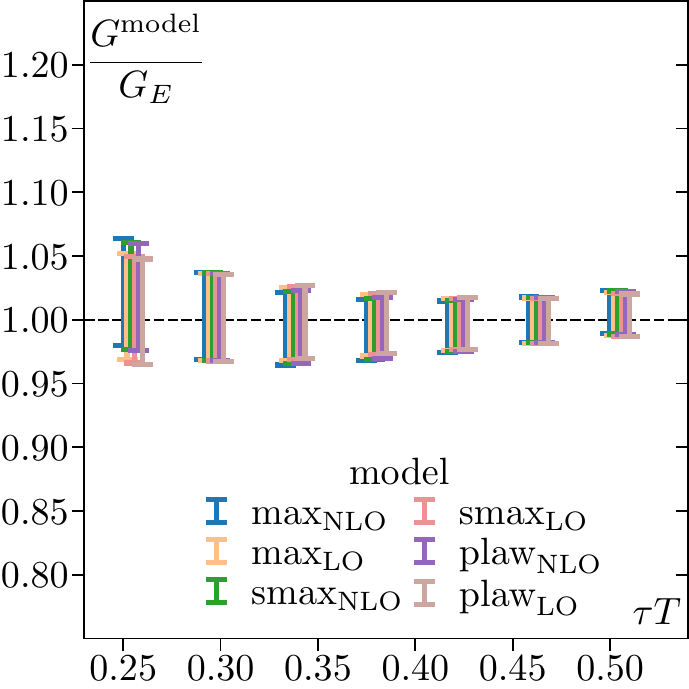}\hfill\null
    \caption{The fitted model correlators divided by the data for $T=195$ MeV (left) and $T=293$ MeV (right).}
    \label{fig:dataofit}
\end{figure*}
\subsection{UV part of the spectral function models}
In order to obtain the heavy quark diffusion coefficient $\kappa$, an Ansatz for
the spectral function $\rho_E$ is needed. The behavior of the spectral function 
at large energies, denoted by $\rho_{\mathrm{UV}}$, is an essential
input for all Ansätze of the spectral function. We use the perturbative LO and NLO results at $T=0$ for $\rho_{\mathrm{UV}}$, multiplied by a free parameter $K$, which accounts for higher-order corrections ($K$-factor).
To evaluate $\rho_{\mathrm{UV}}$,
we need to fix the renormalization scale at which the coupling constant
$g^2$ is defined. At LO, we use the naive choice for the scale $\mu=\max(\mu_T,\omega)$,
where $\mu_T$ is the thermal scale inferred from the high temperature three dimensional
effective theory scale \cite{Kajantie:1997tt}
\begin{equation}
    \mu_T=4 \pi T e^{-\gamma_E-(N_c-8 N_f \ln 2)/(22 N_c-4 N_f)},
\end{equation}
with $N_c$ being the number
of colors and $N_f$ being the number of light quark flavors.
In the context of the chromo-electric spectral function, the use of this scale was
suggested in Ref.~\cite{Burnier:2010rp}.
For the case of interest, $N_c=3$ and $N_f=3$, $\mu_T \simeq 9.08222 T$.
The logarithmic term in the NLO correction sets the natural scale when the NLO result is used.
In this case we use the "optimal" scale that sets the NLO correction to zero 
for large $\omega$ \cite{Burnier:2010rp}:
\begin{equation}
\mu=\mu_{opt}=2 \omega e^{(24 \pi^2-149)N_c+20N_f)/(66 N_c-12 N_f)}.
\end{equation}
For $N_c=3$ and $N_f=3$,
$\mu_{opt} \simeq 14.7427\omega$ is large for all relevant values of $\omega$, so there is no
need to assume that the renormalization scale needs to be temperature dependent.
We use the 5-loop running coupling constant in our analysis as implemented in the  ``AlphasLam'' routine of the RunDec package \cite{Herren:2017osy,Chetyrkin:2000yt} and the value of $\Lambda_{\overline{\mathrm{MS}}}^{N_f=3}=339$ MeV
\cite{FlavourLatticeAveragingGroupFLAG:2021npn,McNeile:2010ji,Chakraborty:2014aca,Ayala:2020odx,Bazavov:2019qoo,Cali:2020hrj,Bruno:2017gxd,PACS-CS:2009zxm,Maltman:2008bx}.

\subsection{Fit results}
\label{supp-mat:SPF}
All of spectral function models considered in this study exhibit two fit parameters: the momentum diffusion coefficient $\kappa/T^3$, and the $K$-factor that multiplies the UV part. The spectral functions corresponding to the fit results for each model are shown in \autoref{fig:spf} for the lowest and the highest temperatures (the intermediate temperatures look similar).
The fits work well as can been seen in \autoref{fig:dataofit}, where we show
the fitted model correlators normalized to the data for different models.

The values of the $K$-factor are shown in \autoref{fig:Kfactor} 
as a function of $T/T_c$ for the $smax$ Ansatz as an example; the results for the other models are again similar.
The value of the $K$-factor is close to one when the LO result is used for $\rho_\mathrm{UV}$, while it is around two when using the NLO result. The situation is somewhat
similar to the one in quenched QCD. The analysis of Refs.~\cite{Francis:2015daa,Altenkort:2020fgs} used the LO result for $\rho_\mathrm{UV}$ and found $K \sim 1$, while the analysis 
of Refs.~\cite{Brambilla:2020siz,Brambilla:2022xbd} that used the NLO result for the UV part of the spectral finds $K \simeq 1.7$ at similar values of $T/T_c$.

We also see from \autoref{fig:spf} that there is a difference between the LO and NLO forms at large $\omega$, but at the same time this does not really seem to effect the slope of $\rho_E/\omega$ in the IR region, meaning that the determination of $\kappa/T^3$ is not affected much by the modeling of the UV part of spectral function, in particular by the value of $K$.
The different values of $K$ for the LO and NLO forms of the spectral functions
are largely due to the difference in the choice of the renormalization scale $\mu$.
For larger values of $\mu$ the $K$ factor for LO form of $\rho^{\mathrm{UV}}$ would be
very different from one.

The $\kappa/T^3$ values obtained from all different
fits are shown in \autoref{fig:kappa_fits} at the lowest and highest
temperature. The temperature dependence of $\kappa/T^3$ is shown in \autoref{fig:asdf}
and \autoref{tab:kappa}. We see that $\kappa/T^3$ decreases with increasing temperature
as also suggest by the weak coupling calculations. We also see that the central values of $\kappa/T^3$ obtained in 2+1-flavor QCD are significantly
larger compared to the ones in quenched QCD, see Refs.~\cite{Altenkort:2020fgs,Brambilla:2020siz,Banerjee:2022gen,Banerjee:2011ra,Francis:2015daa,Brambilla:2022xbd}. We note that quenched results on $\kappa/T^3$ from various lattice groups agree well (cf., Fig.~3 in Ref.~\cite{Banerjee:2022gen}).

\subsection{Details for power-law model}
In the main text we introduced the power-law Ansatz, where $\rho_E(\omega,T)$ is equal to $\rho_\mathrm{IR}(\omega,T)$ for $\omega<\omega_\mathrm{IR}$, and is equal to $\rho_\mathrm{UV}(\omega)$ for $\omega>\omega_\mathrm{UV}$. We need some prior knowledge
to set $\omega_\mathrm{IR}$ and $\omega_\mathrm{UV}$. From the NLO calculations
we know that the thermal corrections to the spectral function are very small for
$\omega>2 \pi T$ \cite{Burnier:2010rp}. Classical calculations in the effective three 
dimensional theory suggest that the spectral function $\rho_E$ is approximately
linear in $\omega$ for $\omega<0.8 T$ \cite{Laine:2009dd}. These calculations are expected
to describe the non-perturbative physics associated with soft chromo-magnetic fields
at high temperatures. The  strong coupling calculations based on
AdS/CFT on the other hand show that the spectral function is approximately linear
up to $\omega \simeq \pi T$ \cite{Gubser:2006nz}. Therefore, for the temperature range of interest, $\omega_\mathrm{IR}$
should not be much smaller than one. We choose $\omega_\mathrm{IR}=T$ in our analysis.
The corresponding fits also work well as one can see from \autoref{fig:dataofit}.
The $K$ factors for these fits are very similar to the ones obtained for 
$max$ and $smax$ forms.

The spectral functions corresponding
to the power law Ansatz are also shown in \autoref{fig:spf} 
at the highest and lowest temperatures. As one can see from the figure, this Ansatz leads to smaller y-intercepts of $\rho_\mathrm{E}/\omega$, and thus smaller values of $\kappa/T^3$.
Choosing larger values of $\omega_\mathrm{IR}$ would result in $\kappa/T^3$ that is closer
to the ones obtain from the $max$ and $smax$ Ans\"atze. As one can see from \autoref{fig:spf},
for the $max$ and $smax$ Ans\"atze, the linear behavior of the spectral function extends
to $\omega \ge \pi T$. Therefore, the  $max$ and $smax$ Ans\"atze correspond to spectral
functions that are more similar to the spectral function in
the strongly coupled theory, while the power law Ansatz with $\omega_\mathrm{IR}=T$ is closer to the spectral 
functions obtained in the weak coupling approach. Thus the $max$ Ansatz and the power-law Ansatz for the spectral function could be interpreted as incorporating features from the strong and weak coupling calculations, respectively.

\subsection{Systematics at nonzero lattice spacing and flow time}

In the main text we saw that the overall shape of the chromo-electric correlator does not change 
much after the continuum and zero flow time extrapolation (cf., \autoref{fig:rawdata} and \ref{fig:contres}). The effects of nonzero lattice spacing and nonzero flow time 
mostly affect the chromo-electric correlation function at small $\tau T$, which are therefore not used in this analysis. Since we cannot perform a continuum extrapolation for the highest temperature ($T=352\,\mathrm{MeV}$), due to a lack of finer lattices, the question arises how much of a systematic error is introduced when performing fits on non-extrapolated correlator data. 
This is further incentivised by considering a generalization of the spectral representation in \autoref{spectral} to nonzero lattice spacing, which implies that the corresponding spectral function has support
only up to some maximal energy $\omega_{max} \sim 1/a$. 
For local (point-like) meson operators this has been shown in Ref.~\cite{Karsch:2003wy}. If one considers extended instead of point-like meson operators, the $\omega_{max}$ of
the corresponding spectral function will decrease \cite{Stickan:2003gh}. However, the spectral functions of the extended meson at $\omega$ significantly smaller than the cutoff will be the same as for the correlator of a local meson \cite{Stickan:2003gh},
if the extent of the meson operators is of the order of few lattice spacings.

Correlators calculated with the gradient flow can be considered as correlators
of extended operators as the gradient flow smears the field within the radius
of $\sqrt{8 \tauf}$.
Based on these considerations we expect that the non-zero lattice spacing and
flow time will mostly affect the large $\omega$ behavior of $\rho_E(\omega,T)$
and only have a small effect on its low $\omega$ behavior and thus the value of $\kappa/T^3$.
For quenched QCD this has been shown in Ref.~\cite{Brambilla:2022xbd}. Therefore,
we performed the above spectral analysis also for the chromo-electric correlator
at $a^{-1}=7.036$ GeV and flow time $\sqrt{8 \tauf}/\tau=0.3$. As expected, 
the fitted spectral function is different in the UV region, which manifest itself in
the smaller value of the $K$ factor as one can see from \autoref{fig:Kfactor}.
On the other hand the values of $\kappa/T^3$ obtained from the analysis of $G_E$
at $a^{-1}=7.036$ GeV and $\sqrt{8 \tauf}/\tau=0.3$ should not change much. As shown in \autoref{fig:asdf}
and \autoref{tab:kappa}, the $\kappa/T^3$ values obtained from the continuum and zero
flow time extrapolated chromo-electric correlator agree well with the ones obtained
at $a^{-1}=7.036$ GeV and $\sqrt{8 \tauf}/\tau=0.3$ within errors, meaning that the statistical
and the systematic uncertainties in $\kappa/T^3$ are much larger than the effects of
non-zero lattice spacing and gradient flow. For this reason we extend our lattice estimate of $\kappa/T^3$ also to $T=352\,\mathrm{MeV}$, which is shown in \autoref{fig:asdf} and \autoref{tab:kappa}.
\begin{figure}[tb]
    \centering
    \includegraphics{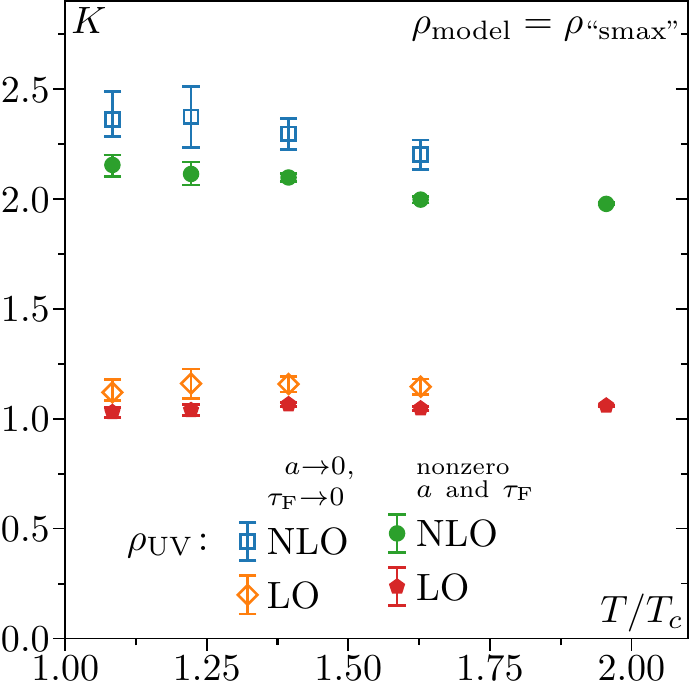}
    \caption{The $K$-factor (i.e., the constant of proportionality between $\rho_\mathrm{UV}$ and $\rho_\mathrm{LO,NLO}$, see text) for the fits of the different temperature data for the \textit{smax} model (see text). The open symbols show the $K$-factor for the fits of the continuum- and flow-time-extrapolated data, while the filled symbols show the $K$-factor for $a^{-1}=7.036\,\mathrm{GeV}$ at 
    $\sqrt{8 \tauf}/\tau=0.3$.}
    \label{fig:Kfactor}
\end{figure}
\begin{figure}[p]
    \hfill%
    \includegraphics{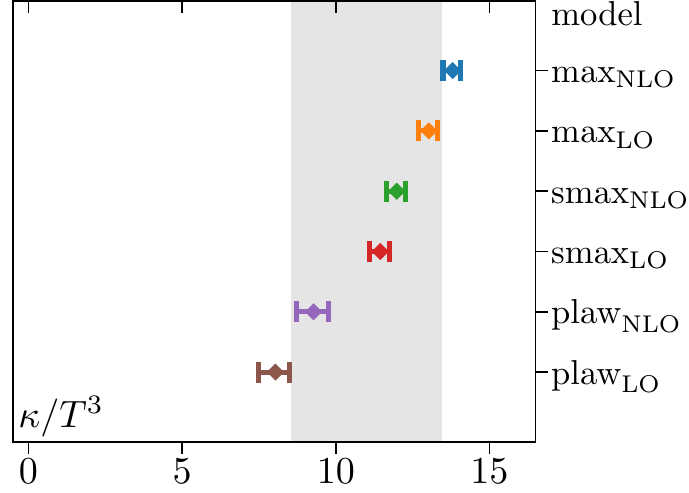}\hspace{2cm}
    \includegraphics{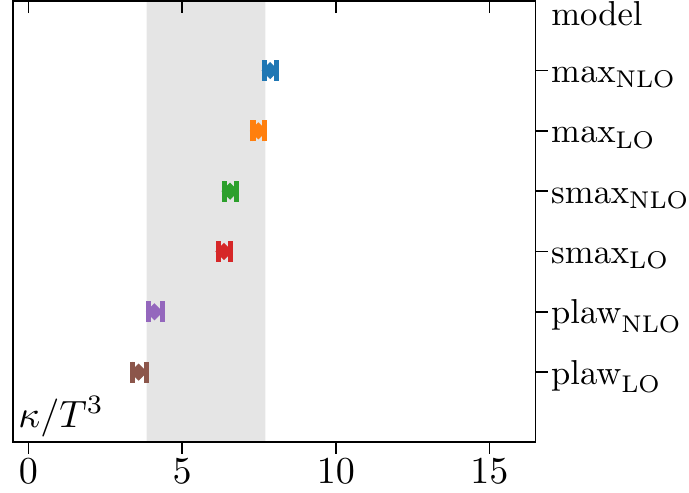}\hfill\null
    \caption{The values of $\kappa/T^3$ obtained using different fit forms
    of the spectral function for 
    $T=195\,\mathrm{MeV}$ (left) and $T=293\,\mathrm{MeV}$ (right) The gray band shows the confidence interval obtained considering
    all possible fits, see main text.}
    \label{fig:kappa_fits}
\end{figure}

\begin{figure}[b]
\begin{minipage}[b]{0.45\textwidth}
    \centering
    \includegraphics{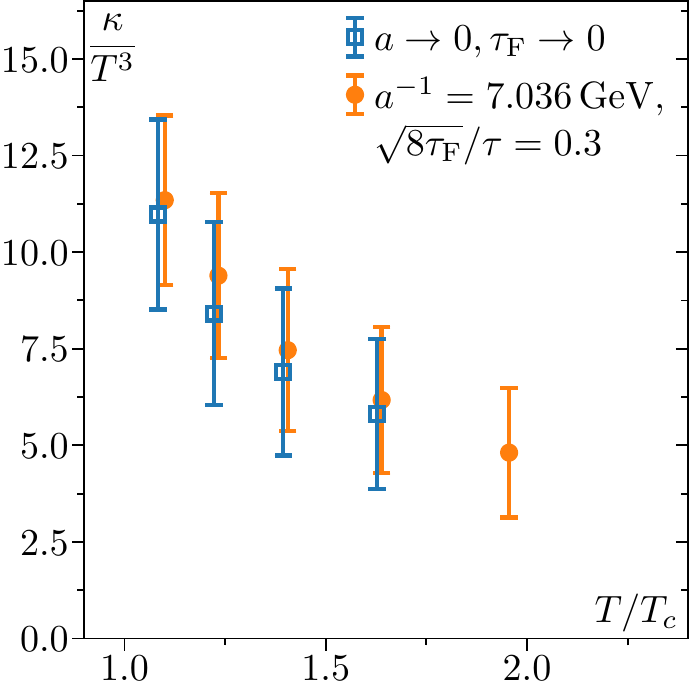}
    \caption{Comparison of the heavy quark momentum diffusion coefficient $\kappa/T^3$ from continuum- ($a\rightarrow 0$) and then flow-time-extrapolated ($\tauf\rightarrow 0$) color-electric correlators and nonzero $(a,\tauf)$ correlators. For better visibility, the nonzero $(a,\tauf)$ results have been slightly offset to the right. The numerical values can be found in \autoref{tab:kappa}.}
    \label{fig:asdf}
\end{minipage}\hfill%
\begin{minipage}[b]{0.45\textwidth}
    \centering
    \begin{tabular}{|r|r|r|}
    \hline
    & \multicolumn{2}{c|}{$\kappa/T^3$} \\
$ T [\mathrm{MeV}] $ & $a \rightarrow 0,\, \tauf \rightarrow 0$  & 
nonzero $a,\, \tauf$ \\
\hline
352 &          -      & $4.8  \pm 1.7$\\
293 & $5.8 \pm 2.0$ & $6.2  \pm 1.9$\\
251 & $6.9 \pm 2.2$ & $7.5  \pm 2.1$\\
220 & $8.4 \pm 2.4$ & $9.4  \pm 2.2$\\
195 & $11.0 \pm 2.5$ & $11.3 \pm 2.2$ \\
\hline
    \end{tabular}
    \captionof{table}{
    The values of $\kappa/T^3$ at different temperatures as shown in \autoref{fig:asdf}. 
    The second column shows the result obtained from the fits of the continuum and zero
    flow time extrapolated results of $G_E$. The third column shows the results for
    $\kappa/T^3$ obtained from the fits of the chromo-electric correlator calculated
    with $a^{-1}=7.036\,\mathrm{GeV}$ at relative flow radius $\sqrt{8 \tauf}/{\tau}=0.3$.\label{tab:kappa}
    }
\end{minipage}
\end{figure}

\end{widetext}

%% file: paper.bbl
\begin{thebibliography}{57}%
\makeatletter
\providecommand \@ifxundefined [1]{%
 \@ifx{#1\undefined}
}%
\providecommand \@ifnum [1]{%
 \ifnum #1\expandafter \@firstoftwo
 \else \expandafter \@secondoftwo
 \fi
}%
\providecommand \@ifx [1]{%
 \ifx #1\expandafter \@firstoftwo
 \else \expandafter \@secondoftwo
 \fi
}%
\providecommand \natexlab [1]{#1}%
\providecommand \enquote  [1]{``#1''}%
\providecommand \bibnamefont  [1]{#1}%
\providecommand \bibfnamefont [1]{#1}%
\providecommand \citenamefont [1]{#1}%
\providecommand \href@noop [0]{\@secondoftwo}%
\providecommand \href [0]{\begingroup \@sanitize@url \@href}%
\providecommand \@href[1]{\@@startlink{#1}\@@href}%
\providecommand \@@href[1]{\endgroup#1\@@endlink}%
\providecommand \@sanitize@url [0]{\catcode `\\12\catcode `\$12\catcode
  `\&12\catcode `\#12\catcode `\^12\catcode `\_12\catcode `\%12\relax}%
\providecommand \@@startlink[1]{}%
\providecommand \@@endlink[0]{}%
\providecommand \url  [0]{\begingroup\@sanitize@url \@url }%
\providecommand \@url [1]{\endgroup\@href {#1}{\urlprefix }}%
\providecommand \urlprefix  [0]{URL }%
\providecommand \Eprint [0]{\href }%
\providecommand \doibase [0]{http://dx.doi.org/}%
\providecommand \selectlanguage [0]{\@gobble}%
\providecommand \bibinfo  [0]{\@secondoftwo}%
\providecommand \bibfield  [0]{\@secondoftwo}%
\providecommand \translation [1]{[#1]}%
\providecommand \BibitemOpen [0]{}%
\providecommand \bibitemStop [0]{}%
\providecommand \bibitemNoStop [0]{.\EOS\space}%
\providecommand \EOS [0]{\spacefactor3000\relax}%
\providecommand \BibitemShut  [1]{\csname bibitem#1\endcsname}%
\let\auto@bib@innerbib\@empty
\bibitem [{\citenamefont {Beraudo}\ \emph {et~al.}(2018)\citenamefont {Beraudo}
  \emph {et~al.}}]{Rapp:2018qla}%
  \BibitemOpen
  \bibfield  {author} {\bibinfo {author} {\bibfnamefont {A.}~\bibnamefont
  {Beraudo}} \emph {et~al.},\ }\href {\doibase 10.1016/j.nuclphysa.2018.09.002}
  {\bibfield  {journal} {\bibinfo  {journal} {Nucl. Phys. A}\ }\textbf
  {\bibinfo {volume} {979}},\ \bibinfo {pages} {21} (\bibinfo {year} {2018})},\
  \Eprint {http://arxiv.org/abs/1803.03824} {arXiv:1803.03824 [nucl-th]}
  \BibitemShut {NoStop}%
\bibitem [{\citenamefont {Dong}\ \emph {et~al.}(2019)\citenamefont {Dong},
  \citenamefont {Lee},\ and\ \citenamefont {Rapp}}]{Dong:2019byy}%
  \BibitemOpen
  \bibfield  {author} {\bibinfo {author} {\bibfnamefont {X.}~\bibnamefont
  {Dong}}, \bibinfo {author} {\bibfnamefont {Y.-J.}\ \bibnamefont {Lee}}, \
  and\ \bibinfo {author} {\bibfnamefont {R.}~\bibnamefont {Rapp}},\ }\href
  {\doibase 10.1146/annurev-nucl-101918-023806} {\bibfield  {journal} {\bibinfo
   {journal} {Ann. Rev. Nucl. Part. Sci.}\ }\textbf {\bibinfo {volume} {69}},\
  \bibinfo {pages} {417} (\bibinfo {year} {2019})},\ \Eprint
  {http://arxiv.org/abs/1903.07709} {arXiv:1903.07709 [nucl-ex]} \BibitemShut
  {NoStop}%
\bibitem [{\citenamefont {He}\ \emph {et~al.}(2022)\citenamefont {He},
  \citenamefont {van Hees},\ and\ \citenamefont {Rapp}}]{He:2022ywp}%
  \BibitemOpen
  \bibfield  {author} {\bibinfo {author} {\bibfnamefont {M.}~\bibnamefont
  {He}}, \bibinfo {author} {\bibfnamefont {H.}~\bibnamefont {van Hees}}, \ and\
  \bibinfo {author} {\bibfnamefont {R.}~\bibnamefont {Rapp}},\ }\href@noop {}
  {\  (\bibinfo {year} {2022})},\ \Eprint {http://arxiv.org/abs/2204.09299}
  {arXiv:2204.09299 [hep-ph]} \BibitemShut {NoStop}%
\bibitem [{\citenamefont {Moore}\ and\ \citenamefont
  {Teaney}(2005)}]{Moore:2004tg}%
  \BibitemOpen
  \bibfield  {author} {\bibinfo {author} {\bibfnamefont {G.~D.}\ \bibnamefont
  {Moore}}\ and\ \bibinfo {author} {\bibfnamefont {D.}~\bibnamefont {Teaney}},\
  }\href {\doibase 10.1103/PhysRevC.71.064904} {\bibfield  {journal} {\bibinfo
  {journal} {Phys. Rev. C}\ }\textbf {\bibinfo {volume} {71}},\ \bibinfo
  {pages} {064904} (\bibinfo {year} {2005})},\ \Eprint
  {http://arxiv.org/abs/hep-ph/0412346} {arXiv:hep-ph/0412346} \BibitemShut
  {NoStop}%
\bibitem [{\citenamefont {Caron-Huot}\ and\ \citenamefont
  {Moore}(2008)}]{CaronHuot:2007gq}%
  \BibitemOpen
  \bibfield  {author} {\bibinfo {author} {\bibfnamefont {S.}~\bibnamefont
  {Caron-Huot}}\ and\ \bibinfo {author} {\bibfnamefont {G.~D.}\ \bibnamefont
  {Moore}},\ }\href {\doibase 10.1103/PhysRevLett.100.052301} {\bibfield
  {journal} {\bibinfo  {journal} {Phys. Rev. Lett.}\ }\textbf {\bibinfo
  {volume} {100}},\ \bibinfo {pages} {052301} (\bibinfo {year} {2008})},\
  \Eprint {http://arxiv.org/abs/0708.4232} {arXiv:0708.4232 [hep-ph]}
  \BibitemShut {NoStop}%
\bibitem [{\citenamefont {Casalderrey-Solana}\ and\ \citenamefont
  {Teaney}(2006)}]{CasalderreySolana:2006rq}%
  \BibitemOpen
  \bibfield  {author} {\bibinfo {author} {\bibfnamefont {J.}~\bibnamefont
  {Casalderrey-Solana}}\ and\ \bibinfo {author} {\bibfnamefont
  {D.}~\bibnamefont {Teaney}},\ }\href {\doibase 10.1103/PhysRevD.74.085012}
  {\bibfield  {journal} {\bibinfo  {journal} {Phys. Rev. D}\ }\textbf {\bibinfo
  {volume} {74}},\ \bibinfo {pages} {085012} (\bibinfo {year} {2006})},\
  \Eprint {http://arxiv.org/abs/hep-ph/0605199} {arXiv:hep-ph/0605199}
  \BibitemShut {NoStop}%
\bibitem [{\citenamefont {Caron-Huot}\ \emph {et~al.}(2009)\citenamefont
  {Caron-Huot}, \citenamefont {Laine},\ and\ \citenamefont
  {Moore}}]{CaronHuot:2009uh}%
  \BibitemOpen
  \bibfield  {author} {\bibinfo {author} {\bibfnamefont {S.}~\bibnamefont
  {Caron-Huot}}, \bibinfo {author} {\bibfnamefont {M.}~\bibnamefont {Laine}}, \
  and\ \bibinfo {author} {\bibfnamefont {G.~D.}\ \bibnamefont {Moore}},\ }\href
  {\doibase 10.1088/1126-6708/2009/04/053} {\bibfield  {journal} {\bibinfo
  {journal} {JHEP}\ }\textbf {\bibinfo {volume} {04}},\ \bibinfo {pages} {053}
  (\bibinfo {year} {2009})},\ \Eprint {http://arxiv.org/abs/0901.1195}
  {arXiv:0901.1195 [hep-lat]} \BibitemShut {NoStop}%
\bibitem [{\citenamefont {Christensen}\ and\ \citenamefont
  {Laine}(2016)}]{Christensen:2016wdo}%
  \BibitemOpen
  \bibfield  {author} {\bibinfo {author} {\bibfnamefont {C.}~\bibnamefont
  {Christensen}}\ and\ \bibinfo {author} {\bibfnamefont {M.}~\bibnamefont
  {Laine}},\ }\href {\doibase 10.1016/j.physletb.2016.02.020} {\bibfield
  {journal} {\bibinfo  {journal} {Phys. Lett. B}\ }\textbf {\bibinfo {volume}
  {755}},\ \bibinfo {pages} {316} (\bibinfo {year} {2016})},\ \Eprint
  {http://arxiv.org/abs/1601.01573} {arXiv:1601.01573 [hep-lat]} \BibitemShut
  {NoStop}%
\bibitem [{\citenamefont {Follana}\ \emph {et~al.}(2007)\citenamefont
  {Follana}, \citenamefont {Mason}, \citenamefont {Davies}, \citenamefont
  {Hornbostel}, \citenamefont {Lepage}, \citenamefont {Shigemitsu},
  \citenamefont {Trottier},\ and\ \citenamefont {Wong}}]{Follana:2006rc}%
  \BibitemOpen
  \bibfield  {author} {\bibinfo {author} {\bibfnamefont {E.}~\bibnamefont
  {Follana}}, \bibinfo {author} {\bibfnamefont {Q.}~\bibnamefont {Mason}},
  \bibinfo {author} {\bibfnamefont {C.}~\bibnamefont {Davies}}, \bibinfo
  {author} {\bibfnamefont {K.}~\bibnamefont {Hornbostel}}, \bibinfo {author}
  {\bibfnamefont {G.~P.}\ \bibnamefont {Lepage}}, \bibinfo {author}
  {\bibfnamefont {J.}~\bibnamefont {Shigemitsu}}, \bibinfo {author}
  {\bibfnamefont {H.}~\bibnamefont {Trottier}}, \ and\ \bibinfo {author}
  {\bibfnamefont {K.}~\bibnamefont {Wong}} (\bibinfo {collaboration} {HPQCD,
  UKQCD}),\ }\href {\doibase 10.1103/PhysRevD.75.054502} {\bibfield  {journal}
  {\bibinfo  {journal} {Phys. Rev.}\ }\textbf {\bibinfo {volume} {D75}},\
  \bibinfo {pages} {054502} (\bibinfo {year} {2007})},\ \Eprint
  {http://arxiv.org/abs/hep-lat/0610092} {arXiv:hep-lat/0610092 [hep-lat]}
  \BibitemShut {NoStop}%
\bibitem [{\citenamefont {Luscher}\ and\ \citenamefont
  {Weisz}(1985{\natexlab{a}})}]{Luscher:1984xn}%
  \BibitemOpen
  \bibfield  {author} {\bibinfo {author} {\bibfnamefont {M.}~\bibnamefont
  {Luscher}}\ and\ \bibinfo {author} {\bibfnamefont {P.}~\bibnamefont
  {Weisz}},\ }\href {\doibase 10.1007/BF01206178} {\bibfield  {journal}
  {\bibinfo  {journal} {Commun. Math. Phys.}\ }\textbf {\bibinfo {volume}
  {97}},\ \bibinfo {pages} {59} (\bibinfo {year} {1985}{\natexlab{a}})},\
  \bibinfo {note} {[Erratum: Commun.Math.Phys. 98, 433 (1985)]}\BibitemShut
  {NoStop}%
\bibitem [{\citenamefont {Luscher}\ and\ \citenamefont
  {Weisz}(1985{\natexlab{b}})}]{Luscher:1985zq}%
  \BibitemOpen
  \bibfield  {author} {\bibinfo {author} {\bibfnamefont {M.}~\bibnamefont
  {Luscher}}\ and\ \bibinfo {author} {\bibfnamefont {P.}~\bibnamefont
  {Weisz}},\ }\href {\doibase 10.1016/0370-2693(85)90966-9} {\bibfield
  {journal} {\bibinfo  {journal} {Phys. Lett. B}\ }\textbf {\bibinfo {volume}
  {158}},\ \bibinfo {pages} {250} (\bibinfo {year}
  {1985}{\natexlab{b}})}\BibitemShut {NoStop}%
\bibitem [{\citenamefont {Bazavov}\ \emph {et~al.}(2014)\citenamefont {Bazavov}
  \emph {et~al.}}]{HotQCD:2014kol}%
  \BibitemOpen
  \bibfield  {author} {\bibinfo {author} {\bibfnamefont {A.}~\bibnamefont
  {Bazavov}} \emph {et~al.} (\bibinfo {collaboration} {HotQCD}),\ }\href
  {\doibase 10.1103/PhysRevD.90.094503} {\bibfield  {journal} {\bibinfo
  {journal} {Phys. Rev. D}\ }\textbf {\bibinfo {volume} {90}},\ \bibinfo
  {pages} {094503} (\bibinfo {year} {2014})},\ \Eprint
  {http://arxiv.org/abs/1407.6387} {arXiv:1407.6387 [hep-lat]} \BibitemShut
  {NoStop}%
\bibitem [{\citenamefont {Bazavov}\ \emph
  {et~al.}(2018{\natexlab{a}})\citenamefont {Bazavov}, \citenamefont
  {Petreczky},\ and\ \citenamefont {Weber}}]{Bazavov:2017dsy}%
  \BibitemOpen
  \bibfield  {author} {\bibinfo {author} {\bibfnamefont {A.}~\bibnamefont
  {Bazavov}}, \bibinfo {author} {\bibfnamefont {P.}~\bibnamefont {Petreczky}},
  \ and\ \bibinfo {author} {\bibfnamefont {J.}~\bibnamefont {Weber}},\ }\href
  {\doibase 10.1103/PhysRevD.97.014510} {\bibfield  {journal} {\bibinfo
  {journal} {Phys. Rev. D}\ }\textbf {\bibinfo {volume} {97}},\ \bibinfo
  {pages} {014510} (\bibinfo {year} {2018}{\natexlab{a}})},\ \Eprint
  {http://arxiv.org/abs/1710.05024} {arXiv:1710.05024 [hep-lat]} \BibitemShut
  {NoStop}%
\bibitem [{sup()}]{supplemental}%
  \BibitemOpen
  \href@noop {} {\bibinfo  {journal} {See Supplemental Material for the
  technical details of this study, which includes
  Refs.~\cite{HotQCD:2014kol,Bazavov:2017dsy,HotQCD:2014kol,Bazavov:2017dsy,Bazavov:2010sb,Bazavov:2010sb,2022-Stendebach,Ramos:2015baa,Altenkort:2020fgs,CaronHuot:2009uh,Eller:2021qpp,Kajantie:1997tt,Burnier:2010rp,Burnier:2010rp,Herren:2017osy,Chetyrkin:2000yt,FlavourLatticeAveragingGroupFLAG:2021npn,McNeile:2010ji,Chakraborty:2014aca,Ayala:2020odx,Bazavov:2019qoo,Cali:2020hrj,Bruno:2017gxd,PACS-CS:2009zxm,Maltman:2008bx,Francis:2015daa,Altenkort:2020fgs,Brambilla:2020siz,Brambilla:2022xbd,Altenkort:2020fgs,Brambilla:2020siz,Banerjee:2022gen,Banerjee:2011ra,Francis:2015daa,Brambilla:2022xbd,Burnier:2010rp,Laine:2009dd,Gubser:2006nz,Karsch:2003wy,Stickan:2003gh,Brambilla:2022xbd,MILC:2010hzw}}\
  }\BibitemShut {NoStop}%
\bibitem [{\citenamefont {Luscher}\ and\ \citenamefont
  {Weisz}(2001)}]{Luscher:2001up}%
  \BibitemOpen
\bibfield  {journal} {  }\bibfield  {author} {\bibinfo {author} {\bibfnamefont
  {M.}~\bibnamefont {Luscher}}\ and\ \bibinfo {author} {\bibfnamefont
  {P.}~\bibnamefont {Weisz}},\ }\href {\doibase 10.1088/1126-6708/2001/09/010}
  {\bibfield  {journal} {\bibinfo  {journal} {JHEP}\ }\textbf {\bibinfo
  {volume} {09}},\ \bibinfo {pages} {010} (\bibinfo {year} {2001})},\ \Eprint
  {http://arxiv.org/abs/hep-lat/0108014} {arXiv:hep-lat/0108014} \BibitemShut
  {NoStop}%
\bibitem [{\citenamefont {Ramos}\ and\ \citenamefont
  {Sint}(2016)}]{Ramos:2015baa}%
  \BibitemOpen
  \bibfield  {author} {\bibinfo {author} {\bibfnamefont {A.}~\bibnamefont
  {Ramos}}\ and\ \bibinfo {author} {\bibfnamefont {S.}~\bibnamefont {Sint}},\
  }\href {\doibase 10.1140/epjc/s10052-015-3831-9} {\bibfield  {journal}
  {\bibinfo  {journal} {Eur. Phys. J. C}\ }\textbf {\bibinfo {volume} {76}},\
  \bibinfo {pages} {15} (\bibinfo {year} {2016})},\ \Eprint
  {http://arxiv.org/abs/1508.05552} {arXiv:1508.05552 [hep-lat]} \BibitemShut
  {NoStop}%
\bibitem [{\citenamefont {L\"uscher}(2010)}]{Luscher:2010iy}%
  \BibitemOpen
  \bibfield  {author} {\bibinfo {author} {\bibfnamefont {M.}~\bibnamefont
  {L\"uscher}},\ }\href {\doibase 10.1007/JHEP08(2010)071} {\bibfield
  {journal} {\bibinfo  {journal} {JHEP}\ }\textbf {\bibinfo {volume} {08}},\
  \bibinfo {pages} {071} (\bibinfo {year} {2010})},\ \bibinfo {note} {[Erratum:
  JHEP 03, 092 (2014)]},\ \Eprint {http://arxiv.org/abs/1006.4518}
  {arXiv:1006.4518 [hep-lat]} \BibitemShut {NoStop}%
\bibitem [{\citenamefont {Altenkort}\ \emph {et~al.}(2021)\citenamefont
  {Altenkort}, \citenamefont {Eller}, \citenamefont {Kaczmarek}, \citenamefont
  {Mazur}, \citenamefont {Moore},\ and\ \citenamefont
  {Shu}}]{Altenkort:2020fgs}%
  \BibitemOpen
  \bibfield  {author} {\bibinfo {author} {\bibfnamefont {L.}~\bibnamefont
  {Altenkort}}, \bibinfo {author} {\bibfnamefont {A.~M.}\ \bibnamefont
  {Eller}}, \bibinfo {author} {\bibfnamefont {O.}~\bibnamefont {Kaczmarek}},
  \bibinfo {author} {\bibfnamefont {L.}~\bibnamefont {Mazur}}, \bibinfo
  {author} {\bibfnamefont {G.~D.}\ \bibnamefont {Moore}}, \ and\ \bibinfo
  {author} {\bibfnamefont {H.-T.}\ \bibnamefont {Shu}},\ }\href {\doibase
  10.1103/PhysRevD.103.014511} {\bibfield  {journal} {\bibinfo  {journal}
  {Phys. Rev. D}\ }\textbf {\bibinfo {volume} {103}},\ \bibinfo {pages}
  {014511} (\bibinfo {year} {2021})},\ \Eprint
  {http://arxiv.org/abs/2009.13553} {arXiv:2009.13553 [hep-lat]} \BibitemShut
  {NoStop}%
\bibitem [{\citenamefont {Brambilla}\ \emph {et~al.}(2023)\citenamefont
  {Brambilla}, \citenamefont {Leino}, \citenamefont {Mayer-Steudte},\ and\
  \citenamefont {Petreczky}}]{Brambilla:2022xbd}%
  \BibitemOpen
  \bibfield  {author} {\bibinfo {author} {\bibfnamefont {N.}~\bibnamefont
  {Brambilla}}, \bibinfo {author} {\bibfnamefont {V.}~\bibnamefont {Leino}},
  \bibinfo {author} {\bibfnamefont {J.}~\bibnamefont {Mayer-Steudte}}, \ and\
  \bibinfo {author} {\bibfnamefont {P.}~\bibnamefont {Petreczky}} (\bibinfo
  {collaboration} {TUMQCD}),\ }\href {\doibase 10.1103/PhysRevD.107.054508}
  {\bibfield  {journal} {\bibinfo  {journal} {Phys. Rev. D}\ }\textbf {\bibinfo
  {volume} {107}},\ \bibinfo {pages} {054508} (\bibinfo {year} {2023})},\
  \Eprint {http://arxiv.org/abs/2206.02861} {arXiv:2206.02861 [hep-lat]}
  \BibitemShut {NoStop}%
\bibitem [{\citenamefont {Eller}\ and\ \citenamefont
  {Moore}(2018)}]{flowlimits}%
  \BibitemOpen
  \bibfield  {author} {\bibinfo {author} {\bibfnamefont {A.~M.}\ \bibnamefont
  {Eller}}\ and\ \bibinfo {author} {\bibfnamefont {G.~D.}\ \bibnamefont
  {Moore}},\ }\href {\doibase 10.1103/PhysRevD.97.114507} {\bibfield  {journal}
  {\bibinfo  {journal} {Phys. Rev. D}\ }\textbf {\bibinfo {volume} {97}},\
  \bibinfo {pages} {114507} (\bibinfo {year} {2018})}\BibitemShut {NoStop}%
\bibitem [{\citenamefont {Brambilla}\ \emph {et~al.}(2020)\citenamefont
  {Brambilla}, \citenamefont {Leino}, \citenamefont {Petreczky},\ and\
  \citenamefont {Vairo}}]{Brambilla:2020siz}%
  \BibitemOpen
  \bibfield  {author} {\bibinfo {author} {\bibfnamefont {N.}~\bibnamefont
  {Brambilla}}, \bibinfo {author} {\bibfnamefont {V.}~\bibnamefont {Leino}},
  \bibinfo {author} {\bibfnamefont {P.}~\bibnamefont {Petreczky}}, \ and\
  \bibinfo {author} {\bibfnamefont {A.}~\bibnamefont {Vairo}},\ }\href
  {\doibase 10.1103/PhysRevD.102.074503} {\bibfield  {journal} {\bibinfo
  {journal} {Phys. Rev. D}\ }\textbf {\bibinfo {volume} {102}},\ \bibinfo
  {pages} {074503} (\bibinfo {year} {2020})},\ \Eprint
  {http://arxiv.org/abs/2007.10078} {arXiv:2007.10078 [hep-lat]} \BibitemShut
  {NoStop}%
\bibitem [{\citenamefont {Eller}(2021)}]{Eller:2021qpp}%
  \BibitemOpen
  \bibfield  {author} {\bibinfo {author} {\bibfnamefont {A.~M.}\ \bibnamefont
  {Eller}},\ }\emph {\bibinfo {title} {{The Color-Electric Field Correlator
  under Gradient Flow at next-to-leading Order in Quantum Chromodynamics}}},\
  \href {\doibase 10.26083/tuprints-00017610} {Ph.D. thesis},\ \bibinfo
  {school} {Tech. U., Dortmund (main), Darmstadt, Tech. Hochsch.} (\bibinfo
  {year} {2021})\BibitemShut {NoStop}%
\bibitem [{\citenamefont {Burnier}\ \emph {et~al.}(2010)\citenamefont
  {Burnier}, \citenamefont {Laine}, \citenamefont {Langelage},\ and\
  \citenamefont {Mether}}]{Burnier:2010rp}%
  \BibitemOpen
  \bibfield  {author} {\bibinfo {author} {\bibfnamefont {Y.}~\bibnamefont
  {Burnier}}, \bibinfo {author} {\bibfnamefont {M.}~\bibnamefont {Laine}},
  \bibinfo {author} {\bibfnamefont {J.}~\bibnamefont {Langelage}}, \ and\
  \bibinfo {author} {\bibfnamefont {L.}~\bibnamefont {Mether}},\ }\href
  {\doibase 10.1007/JHEP08(2010)094} {\bibfield  {journal} {\bibinfo  {journal}
  {JHEP}\ }\textbf {\bibinfo {volume} {08}},\ \bibinfo {pages} {094} (\bibinfo
  {year} {2010})},\ \Eprint {http://arxiv.org/abs/1006.0867} {arXiv:1006.0867
  [hep-ph]} \BibitemShut {NoStop}%
\bibitem [{\citenamefont {Laine}\ \emph {et~al.}(2009)\citenamefont {Laine},
  \citenamefont {Moore}, \citenamefont {Philipsen},\ and\ \citenamefont
  {Tassler}}]{Laine:2009dd}%
  \BibitemOpen
  \bibfield  {author} {\bibinfo {author} {\bibfnamefont {M.}~\bibnamefont
  {Laine}}, \bibinfo {author} {\bibfnamefont {G.~D.}\ \bibnamefont {Moore}},
  \bibinfo {author} {\bibfnamefont {O.}~\bibnamefont {Philipsen}}, \ and\
  \bibinfo {author} {\bibfnamefont {M.}~\bibnamefont {Tassler}},\ }\href
  {\doibase 10.1088/1126-6708/2009/05/014} {\bibfield  {journal} {\bibinfo
  {journal} {JHEP}\ }\textbf {\bibinfo {volume} {05}},\ \bibinfo {pages} {014}
  (\bibinfo {year} {2009})},\ \Eprint {http://arxiv.org/abs/0902.2856}
  {arXiv:0902.2856 [hep-ph]} \BibitemShut {NoStop}%
\bibitem [{\citenamefont {Gubser}(2008)}]{Gubser:2006nz}%
  \BibitemOpen
  \bibfield  {author} {\bibinfo {author} {\bibfnamefont {S.~S.}\ \bibnamefont
  {Gubser}},\ }\href {\doibase 10.1016/j.nuclphysb.2007.09.017} {\bibfield
  {journal} {\bibinfo  {journal} {Nucl. Phys. B}\ }\textbf {\bibinfo {volume}
  {790}},\ \bibinfo {pages} {175} (\bibinfo {year} {2008})},\ \Eprint
  {http://arxiv.org/abs/hep-th/0612143} {arXiv:hep-th/0612143} \BibitemShut
  {NoStop}%
\bibitem [{\citenamefont {Francis}\ \emph {et~al.}(2015)\citenamefont
  {Francis}, \citenamefont {Kaczmarek}, \citenamefont {Laine}, \citenamefont
  {Neuhaus},\ and\ \citenamefont {Ohno}}]{Francis:2015daa}%
  \BibitemOpen
  \bibfield  {author} {\bibinfo {author} {\bibfnamefont {A.}~\bibnamefont
  {Francis}}, \bibinfo {author} {\bibfnamefont {O.}~\bibnamefont {Kaczmarek}},
  \bibinfo {author} {\bibfnamefont {M.}~\bibnamefont {Laine}}, \bibinfo
  {author} {\bibfnamefont {T.}~\bibnamefont {Neuhaus}}, \ and\ \bibinfo
  {author} {\bibfnamefont {H.}~\bibnamefont {Ohno}},\ }\href {\doibase
  10.1103/PhysRevD.92.116003} {\bibfield  {journal} {\bibinfo  {journal} {Phys.
  Rev. D}\ }\textbf {\bibinfo {volume} {92}},\ \bibinfo {pages} {116003}
  (\bibinfo {year} {2015})},\ \Eprint {http://arxiv.org/abs/1508.04543}
  {arXiv:1508.04543 [hep-lat]} \BibitemShut {NoStop}%
\bibitem [{\citenamefont {Caron-Huot}(2009)}]{Caron-Huot:2009ypo}%
  \BibitemOpen
  \bibfield  {author} {\bibinfo {author} {\bibfnamefont {S.}~\bibnamefont
  {Caron-Huot}},\ }\href {\doibase 10.1103/PhysRevD.79.125009} {\bibfield
  {journal} {\bibinfo  {journal} {Phys. Rev. D}\ }\textbf {\bibinfo {volume}
  {79}},\ \bibinfo {pages} {125009} (\bibinfo {year} {2009})},\ \Eprint
  {http://arxiv.org/abs/0903.3958} {arXiv:0903.3958 [hep-ph]} \BibitemShut
  {NoStop}%
\bibitem [{\citenamefont {Banerjee}\ \emph
  {et~al.}(2022{\natexlab{a}})\citenamefont {Banerjee}, \citenamefont {Gavai},
  \citenamefont {Datta},\ and\ \citenamefont {Majumdar}}]{Banerjee:2022gen}%
  \BibitemOpen
  \bibfield  {author} {\bibinfo {author} {\bibfnamefont {D.}~\bibnamefont
  {Banerjee}}, \bibinfo {author} {\bibfnamefont {R.}~\bibnamefont {Gavai}},
  \bibinfo {author} {\bibfnamefont {S.}~\bibnamefont {Datta}}, \ and\ \bibinfo
  {author} {\bibfnamefont {P.}~\bibnamefont {Majumdar}},\ }\href@noop {} {\
  (\bibinfo {year} {2022}{\natexlab{a}})},\ \Eprint
  {http://arxiv.org/abs/2206.15471} {arXiv:2206.15471 [hep-ph]} \BibitemShut
  {NoStop}%
\bibitem [{\citenamefont {Liu}\ and\ \citenamefont {Rapp}(2020)}]{Liu:2016ysz}%
  \BibitemOpen
  \bibfield  {author} {\bibinfo {author} {\bibfnamefont {S.~Y.~F.}\
  \bibnamefont {Liu}}\ and\ \bibinfo {author} {\bibfnamefont {R.}~\bibnamefont
  {Rapp}},\ }\href {\doibase 10.1140/epja/s10050-020-00024-z} {\bibfield
  {journal} {\bibinfo  {journal} {Eur. Phys. J. A}\ }\textbf {\bibinfo {volume}
  {56}},\ \bibinfo {pages} {44} (\bibinfo {year} {2020})},\ \Eprint
  {http://arxiv.org/abs/1612.09138} {arXiv:1612.09138 [nucl-th]} \BibitemShut
  {NoStop}%
\bibitem [{\citenamefont {Liu}\ and\ \citenamefont {Rapp}(2018)}]{Liu:2017qah}%
  \BibitemOpen
  \bibfield  {author} {\bibinfo {author} {\bibfnamefont {S.~Y.~F.}\
  \bibnamefont {Liu}}\ and\ \bibinfo {author} {\bibfnamefont {R.}~\bibnamefont
  {Rapp}},\ }\href {\doibase 10.1103/PhysRevC.97.034918} {\bibfield  {journal}
  {\bibinfo  {journal} {Phys. Rev. C}\ }\textbf {\bibinfo {volume} {97}},\
  \bibinfo {pages} {034918} (\bibinfo {year} {2018})},\ \Eprint
  {http://arxiv.org/abs/1711.03282} {arXiv:1711.03282 [nucl-th]} \BibitemShut
  {NoStop}%
\bibitem [{\citenamefont {Xu}\ \emph {et~al.}(2018)\citenamefont {Xu},
  \citenamefont {Bernhard}, \citenamefont {Bass}, \citenamefont {Nahrgang},\
  and\ \citenamefont {Cao}}]{Xu:2017obm}%
  \BibitemOpen
  \bibfield  {author} {\bibinfo {author} {\bibfnamefont {Y.}~\bibnamefont
  {Xu}}, \bibinfo {author} {\bibfnamefont {J.~E.}\ \bibnamefont {Bernhard}},
  \bibinfo {author} {\bibfnamefont {S.~A.}\ \bibnamefont {Bass}}, \bibinfo
  {author} {\bibfnamefont {M.}~\bibnamefont {Nahrgang}}, \ and\ \bibinfo
  {author} {\bibfnamefont {S.}~\bibnamefont {Cao}},\ }\href {\doibase
  10.1103/PhysRevC.97.014907} {\bibfield  {journal} {\bibinfo  {journal} {Phys.
  Rev. C}\ }\textbf {\bibinfo {volume} {97}},\ \bibinfo {pages} {014907}
  (\bibinfo {year} {2018})},\ \Eprint {http://arxiv.org/abs/1710.00807}
  {arXiv:1710.00807 [nucl-th]} \BibitemShut {NoStop}%
\bibitem [{\citenamefont {Acharya}\ \emph {et~al.}(2022)\citenamefont {Acharya}
  \emph {et~al.}}]{ALICE:2021rxa}%
  \BibitemOpen
  \bibfield  {author} {\bibinfo {author} {\bibfnamefont {S.}~\bibnamefont
  {Acharya}} \emph {et~al.} (\bibinfo {collaboration} {ALICE}),\ }\href
  {\doibase 10.1007/JHEP01(2022)174} {\bibfield  {journal} {\bibinfo  {journal}
  {JHEP}\ }\textbf {\bibinfo {volume} {01}},\ \bibinfo {pages} {174} (\bibinfo
  {year} {2022})},\ \Eprint {http://arxiv.org/abs/2110.09420} {arXiv:2110.09420
  [nucl-ex]} \BibitemShut {NoStop}%
\bibitem [{\citenamefont {Andreev}(2018)}]{Andreev:2017bvr}%
  \BibitemOpen
  \bibfield  {author} {\bibinfo {author} {\bibfnamefont {O.}~\bibnamefont
  {Andreev}},\ }\href {\doibase 10.1142/S0217732318500414} {\bibfield
  {journal} {\bibinfo  {journal} {Mod. Phys. Lett. A}\ }\textbf {\bibinfo
  {volume} {33}},\ \bibinfo {pages} {1850041} (\bibinfo {year} {2018})},\
  \Eprint {http://arxiv.org/abs/1707.05045} {arXiv:1707.05045 [hep-ph]}
  \BibitemShut {NoStop}%
\bibitem [{\citenamefont {Bazavov}\ and\ \citenamefont
  {Petreczky}(2011)}]{Bazavov:2010bx}%
  \BibitemOpen
  \bibfield  {author} {\bibinfo {author} {\bibfnamefont {A.}~\bibnamefont
  {Bazavov}}\ and\ \bibinfo {author} {\bibfnamefont {P.}~\bibnamefont
  {Petreczky}} (\bibinfo {collaboration} {HotQCD}),\ }\href {\doibase
  10.1134/S1547477111080024} {\bibfield  {journal} {\bibinfo  {journal} {Phys.
  Part. Nucl. Lett.}\ }\textbf {\bibinfo {volume} {8}},\ \bibinfo {pages} {860}
  (\bibinfo {year} {2011})},\ \Eprint {http://arxiv.org/abs/1009.4914}
  {arXiv:1009.4914 [hep-lat]} \BibitemShut {NoStop}%
\bibitem [{\citenamefont {Bazavov}\ and\ \citenamefont
  {Petreczky}(2010)}]{Bazavov:2010sb}%
  \BibitemOpen
  \bibfield  {author} {\bibinfo {author} {\bibfnamefont {A.}~\bibnamefont
  {Bazavov}}\ and\ \bibinfo {author} {\bibfnamefont {P.}~\bibnamefont
  {Petreczky}} (\bibinfo {collaboration} {HotQCD}),\ }\href {\doibase
  10.1088/1742-6596/230/1/012014} {\bibfield  {journal} {\bibinfo  {journal}
  {J. Phys. Conf. Ser.}\ }\textbf {\bibinfo {volume} {230}},\ \bibinfo {pages}
  {012014} (\bibinfo {year} {2010})},\ \Eprint {http://arxiv.org/abs/1005.1131}
  {arXiv:1005.1131 [hep-lat]} \BibitemShut {NoStop}%
\bibitem [{\citenamefont {Bazavov}\ \emph
  {et~al.}(2018{\natexlab{b}})\citenamefont {Bazavov}, \citenamefont
  {Brambilla}, \citenamefont {Petreczky}, \citenamefont {Vairo},\ and\
  \citenamefont {Weber}}]{Bazavov:2018wmo}%
  \BibitemOpen
  \bibfield  {author} {\bibinfo {author} {\bibfnamefont {A.}~\bibnamefont
  {Bazavov}}, \bibinfo {author} {\bibfnamefont {N.}~\bibnamefont {Brambilla}},
  \bibinfo {author} {\bibfnamefont {P.}~\bibnamefont {Petreczky}}, \bibinfo
  {author} {\bibfnamefont {A.}~\bibnamefont {Vairo}}, \ and\ \bibinfo {author}
  {\bibfnamefont {J.~H.}\ \bibnamefont {Weber}} (\bibinfo {collaboration}
  {TUMQCD}),\ }\href {\doibase 10.1103/PhysRevD.98.054511} {\bibfield
  {journal} {\bibinfo  {journal} {Phys. Rev. D}\ }\textbf {\bibinfo {volume}
  {98}},\ \bibinfo {pages} {054511} (\bibinfo {year} {2018}{\natexlab{b}})},\
  \Eprint {http://arxiv.org/abs/1804.10600} {arXiv:1804.10600 [hep-lat]}
  \BibitemShut {NoStop}%
\bibitem [{\citenamefont {Banerjee}\ \emph
  {et~al.}(2022{\natexlab{b}})\citenamefont {Banerjee}, \citenamefont {Datta},\
  and\ \citenamefont {Laine}}]{Banerjee:2022uge}%
  \BibitemOpen
  \bibfield  {author} {\bibinfo {author} {\bibfnamefont {D.}~\bibnamefont
  {Banerjee}}, \bibinfo {author} {\bibfnamefont {S.}~\bibnamefont {Datta}}, \
  and\ \bibinfo {author} {\bibfnamefont {M.}~\bibnamefont {Laine}},\ }\href
  {\doibase 10.1007/JHEP08(2022)128} {\bibfield  {journal} {\bibinfo  {journal}
  {JHEP}\ }\textbf {\bibinfo {volume} {08}},\ \bibinfo {pages} {128} (\bibinfo
  {year} {2022}{\natexlab{b}})},\ \Eprint {http://arxiv.org/abs/2204.14075}
  {arXiv:2204.14075 [hep-lat]} \BibitemShut {NoStop}%
\bibitem [{\citenamefont {Altenkort}\ \emph {et~al.}(2023)\citenamefont
  {Altenkort}, \citenamefont {Kaczmarek}, \citenamefont {R.}, \citenamefont
  {S.}, \citenamefont {P.}, \citenamefont {Shu},\ and\ \citenamefont
  {Stendebach}}]{datapublication}%
  \BibitemOpen
  \bibfield  {author} {\bibinfo {author} {\bibfnamefont {L.}~\bibnamefont
  {Altenkort}}, \bibinfo {author} {\bibfnamefont {O.}~\bibnamefont
  {Kaczmarek}}, \bibinfo {author} {\bibfnamefont {L.}~\bibnamefont {R.}},
  \bibinfo {author} {\bibfnamefont {M.}~\bibnamefont {S.}}, \bibinfo {author}
  {\bibfnamefont {P.}~\bibnamefont {P.}}, \bibinfo {author} {\bibfnamefont
  {H.-T.}\ \bibnamefont {Shu}}, \ and\ \bibinfo {author} {\bibfnamefont
  {S.}~\bibnamefont {Stendebach}},\ }\href {\doibase 10.4119/unibi/2979080}
  {\bibfield  {journal} {\bibinfo  {journal} {Bielefeld University}\ }
  (\bibinfo {year} {2023}),\ 10.4119/unibi/2979080}\BibitemShut {NoStop}%
\bibitem [{\citenamefont {Mazur}(2021)}]{Mazur:2021zgi}%
  \BibitemOpen
  \bibfield  {author} {\bibinfo {author} {\bibfnamefont {L.}~\bibnamefont
  {Mazur}},\ }\emph {\bibinfo {title} {{Topological Aspects in Lattice QCD}}},\
  \href {\doibase 10.4119/unibi/2956493} {Ph.D. thesis},\ \bibinfo  {school}
  {Bielefeld U.} (\bibinfo {year} {2021})\BibitemShut {NoStop}%
\bibitem [{\citenamefont {Bollweg}\ \emph {et~al.}(2022)\citenamefont
  {Bollweg}, \citenamefont {Altenkort}, \citenamefont {Clarke}, \citenamefont
  {Kaczmarek}, \citenamefont {Mazur}, \citenamefont {Schmidt}, \citenamefont
  {Scior},\ and\ \citenamefont {Shu}}]{Bollweg:2021cvl}%
  \BibitemOpen
  \bibfield  {author} {\bibinfo {author} {\bibfnamefont {D.}~\bibnamefont
  {Bollweg}}, \bibinfo {author} {\bibfnamefont {L.}~\bibnamefont {Altenkort}},
  \bibinfo {author} {\bibfnamefont {D.~A.}\ \bibnamefont {Clarke}}, \bibinfo
  {author} {\bibfnamefont {O.}~\bibnamefont {Kaczmarek}}, \bibinfo {author}
  {\bibfnamefont {L.}~\bibnamefont {Mazur}}, \bibinfo {author} {\bibfnamefont
  {C.}~\bibnamefont {Schmidt}}, \bibinfo {author} {\bibfnamefont
  {P.}~\bibnamefont {Scior}}, \ and\ \bibinfo {author} {\bibfnamefont {H.-T.}\
  \bibnamefont {Shu}},\ }\href {\doibase 10.22323/1.396.0196} {\bibfield
  {journal} {\bibinfo  {journal} {PoS}\ }\textbf {\bibinfo {volume}
  {LATTICE2021}},\ \bibinfo {pages} {196} (\bibinfo {year} {2022})},\ \Eprint
  {http://arxiv.org/abs/2111.10354} {arXiv:2111.10354 [hep-lat]} \BibitemShut
  {NoStop}%
\bibitem [{\citenamefont {Bazavov}\ \emph {et~al.}(2010)\citenamefont {Bazavov}
  \emph {et~al.}}]{MILC:2010hzw}%
  \BibitemOpen
  \bibfield  {author} {\bibinfo {author} {\bibfnamefont {A.}~\bibnamefont
  {Bazavov}} \emph {et~al.} (\bibinfo {collaboration} {MILC}),\ }\href
  {\doibase 10.22323/1.105.0074} {\bibfield  {journal} {\bibinfo  {journal}
  {PoS}\ }\textbf {\bibinfo {volume} {LATTICE2010}},\ \bibinfo {pages} {074}
  (\bibinfo {year} {2010})},\ \Eprint {http://arxiv.org/abs/1012.0868}
  {arXiv:1012.0868 [hep-lat]} \BibitemShut {NoStop}%
\bibitem [{\citenamefont {Stendebach}(2022)}]{2022-Stendebach}%
  \BibitemOpen
  \bibfield  {author} {\bibinfo {author} {\bibfnamefont {S.}~\bibnamefont
  {Stendebach}},\ }\emph {\bibinfo {title} {{Perturbative analysis of operators
  under improved gradient flow in lattice QCD}}},\ \href {\doibase
  https://doi.org/10.26083/tuprints-00023185} {Master's thesis},\ \bibinfo
  {school} {Technische Universität Darmstadt} (\bibinfo {year}
  {2022})\BibitemShut {NoStop}%
\bibitem [{\citenamefont {Kajantie}\ \emph {et~al.}(1997)\citenamefont
  {Kajantie}, \citenamefont {Laine}, \citenamefont {Rummukainen},\ and\
  \citenamefont {Shaposhnikov}}]{Kajantie:1997tt}%
  \BibitemOpen
  \bibfield  {author} {\bibinfo {author} {\bibfnamefont {K.}~\bibnamefont
  {Kajantie}}, \bibinfo {author} {\bibfnamefont {M.}~\bibnamefont {Laine}},
  \bibinfo {author} {\bibfnamefont {K.}~\bibnamefont {Rummukainen}}, \ and\
  \bibinfo {author} {\bibfnamefont {M.~E.}\ \bibnamefont {Shaposhnikov}},\
  }\href {\doibase 10.1016/S0550-3213(97)00425-2} {\bibfield  {journal}
  {\bibinfo  {journal} {Nucl. Phys. B}\ }\textbf {\bibinfo {volume} {503}},\
  \bibinfo {pages} {357} (\bibinfo {year} {1997})},\ \Eprint
  {http://arxiv.org/abs/hep-ph/9704416} {arXiv:hep-ph/9704416} \BibitemShut
  {NoStop}%
\bibitem [{\citenamefont {Herren}\ and\ \citenamefont
  {Steinhauser}(2018)}]{Herren:2017osy}%
  \BibitemOpen
  \bibfield  {author} {\bibinfo {author} {\bibfnamefont {F.}~\bibnamefont
  {Herren}}\ and\ \bibinfo {author} {\bibfnamefont {M.}~\bibnamefont
  {Steinhauser}},\ }\href {\doibase 10.1016/j.cpc.2017.11.014} {\bibfield
  {journal} {\bibinfo  {journal} {Comput. Phys. Commun.}\ }\textbf {\bibinfo
  {volume} {224}},\ \bibinfo {pages} {333} (\bibinfo {year} {2018})},\ \Eprint
  {http://arxiv.org/abs/1703.03751} {arXiv:1703.03751 [hep-ph]} \BibitemShut
  {NoStop}%
\bibitem [{\citenamefont {Chetyrkin}\ \emph {et~al.}(2000)\citenamefont
  {Chetyrkin}, \citenamefont {Kuhn},\ and\ \citenamefont
  {Steinhauser}}]{Chetyrkin:2000yt}%
  \BibitemOpen
  \bibfield  {author} {\bibinfo {author} {\bibfnamefont {K.~G.}\ \bibnamefont
  {Chetyrkin}}, \bibinfo {author} {\bibfnamefont {J.~H.}\ \bibnamefont {Kuhn}},
  \ and\ \bibinfo {author} {\bibfnamefont {M.}~\bibnamefont {Steinhauser}},\
  }\href {\doibase 10.1016/S0010-4655(00)00155-7} {\bibfield  {journal}
  {\bibinfo  {journal} {Comput. Phys. Commun.}\ }\textbf {\bibinfo {volume}
  {133}},\ \bibinfo {pages} {43} (\bibinfo {year} {2000})},\ \Eprint
  {http://arxiv.org/abs/hep-ph/0004189} {arXiv:hep-ph/0004189} \BibitemShut
  {NoStop}%
\bibitem [{\citenamefont {Aoki}\ \emph {et~al.}(2022)\citenamefont {Aoki} \emph
  {et~al.}}]{FlavourLatticeAveragingGroupFLAG:2021npn}%
  \BibitemOpen
  \bibfield  {author} {\bibinfo {author} {\bibfnamefont {Y.}~\bibnamefont
  {Aoki}} \emph {et~al.} (\bibinfo {collaboration} {Flavour Lattice Averaging
  Group (FLAG)}),\ }\href {\doibase 10.1140/epjc/s10052-022-10536-1} {\bibfield
   {journal} {\bibinfo  {journal} {Eur. Phys. J. C}\ }\textbf {\bibinfo
  {volume} {82}},\ \bibinfo {pages} {869} (\bibinfo {year} {2022})},\ \Eprint
  {http://arxiv.org/abs/2111.09849} {arXiv:2111.09849 [hep-lat]} \BibitemShut
  {NoStop}%
\bibitem [{\citenamefont {McNeile}\ \emph {et~al.}(2010)\citenamefont
  {McNeile}, \citenamefont {Davies}, \citenamefont {Follana}, \citenamefont
  {Hornbostel},\ and\ \citenamefont {Lepage}}]{McNeile:2010ji}%
  \BibitemOpen
  \bibfield  {author} {\bibinfo {author} {\bibfnamefont {C.}~\bibnamefont
  {McNeile}}, \bibinfo {author} {\bibfnamefont {C.~T.~H.}\ \bibnamefont
  {Davies}}, \bibinfo {author} {\bibfnamefont {E.}~\bibnamefont {Follana}},
  \bibinfo {author} {\bibfnamefont {K.}~\bibnamefont {Hornbostel}}, \ and\
  \bibinfo {author} {\bibfnamefont {G.~P.}\ \bibnamefont {Lepage}},\ }\href
  {\doibase 10.1103/PhysRevD.82.034512} {\bibfield  {journal} {\bibinfo
  {journal} {Phys. Rev. D}\ }\textbf {\bibinfo {volume} {82}},\ \bibinfo
  {pages} {034512} (\bibinfo {year} {2010})},\ \Eprint
  {http://arxiv.org/abs/1004.4285} {arXiv:1004.4285 [hep-lat]} \BibitemShut
  {NoStop}%
\bibitem [{\citenamefont {Chakraborty}\ \emph {et~al.}(2015)\citenamefont
  {Chakraborty}, \citenamefont {Davies}, \citenamefont {Galloway},
  \citenamefont {Knecht}, \citenamefont {Koponen}, \citenamefont {Donald},
  \citenamefont {Dowdall}, \citenamefont {Lepage},\ and\ \citenamefont
  {McNeile}}]{Chakraborty:2014aca}%
  \BibitemOpen
  \bibfield  {author} {\bibinfo {author} {\bibfnamefont {B.}~\bibnamefont
  {Chakraborty}}, \bibinfo {author} {\bibfnamefont {C.~T.~H.}\ \bibnamefont
  {Davies}}, \bibinfo {author} {\bibfnamefont {B.}~\bibnamefont {Galloway}},
  \bibinfo {author} {\bibfnamefont {P.}~\bibnamefont {Knecht}}, \bibinfo
  {author} {\bibfnamefont {J.}~\bibnamefont {Koponen}}, \bibinfo {author}
  {\bibfnamefont {G.~C.}\ \bibnamefont {Donald}}, \bibinfo {author}
  {\bibfnamefont {R.~J.}\ \bibnamefont {Dowdall}}, \bibinfo {author}
  {\bibfnamefont {G.~P.}\ \bibnamefont {Lepage}}, \ and\ \bibinfo {author}
  {\bibfnamefont {C.}~\bibnamefont {McNeile}},\ }\href {\doibase
  10.1103/PhysRevD.91.054508} {\bibfield  {journal} {\bibinfo  {journal} {Phys.
  Rev. D}\ }\textbf {\bibinfo {volume} {91}},\ \bibinfo {pages} {054508}
  (\bibinfo {year} {2015})},\ \Eprint {http://arxiv.org/abs/1408.4169}
  {arXiv:1408.4169 [hep-lat]} \BibitemShut {NoStop}%
\bibitem [{\citenamefont {Ayala}\ \emph {et~al.}(2020)\citenamefont {Ayala},
  \citenamefont {Lobregat},\ and\ \citenamefont {Pineda}}]{Ayala:2020odx}%
  \BibitemOpen
  \bibfield  {author} {\bibinfo {author} {\bibfnamefont {C.}~\bibnamefont
  {Ayala}}, \bibinfo {author} {\bibfnamefont {X.}~\bibnamefont {Lobregat}}, \
  and\ \bibinfo {author} {\bibfnamefont {A.}~\bibnamefont {Pineda}},\ }\href
  {\doibase 10.1007/JHEP09(2020)016} {\bibfield  {journal} {\bibinfo  {journal}
  {JHEP}\ }\textbf {\bibinfo {volume} {09}},\ \bibinfo {pages} {016} (\bibinfo
  {year} {2020})},\ \Eprint {http://arxiv.org/abs/2005.12301} {arXiv:2005.12301
  [hep-ph]} \BibitemShut {NoStop}%
\bibitem [{\citenamefont {Bazavov}\ \emph {et~al.}(2019)\citenamefont
  {Bazavov}, \citenamefont {Brambilla}, \citenamefont {Garcia~i Tormo},
  \citenamefont {Petreczky}, \citenamefont {Soto}, \citenamefont {Vairo},\ and\
  \citenamefont {Weber}}]{Bazavov:2019qoo}%
  \BibitemOpen
  \bibfield  {author} {\bibinfo {author} {\bibfnamefont {A.}~\bibnamefont
  {Bazavov}}, \bibinfo {author} {\bibfnamefont {N.}~\bibnamefont {Brambilla}},
  \bibinfo {author} {\bibfnamefont {X.}~\bibnamefont {Garcia~i Tormo}},
  \bibinfo {author} {\bibfnamefont {P.}~\bibnamefont {Petreczky}}, \bibinfo
  {author} {\bibfnamefont {J.}~\bibnamefont {Soto}}, \bibinfo {author}
  {\bibfnamefont {A.}~\bibnamefont {Vairo}}, \ and\ \bibinfo {author}
  {\bibfnamefont {J.~H.}\ \bibnamefont {Weber}} (\bibinfo {collaboration}
  {TUMQCD}),\ }\href {\doibase 10.1103/PhysRevD.100.114511} {\bibfield
  {journal} {\bibinfo  {journal} {Phys. Rev. D}\ }\textbf {\bibinfo {volume}
  {100}},\ \bibinfo {pages} {114511} (\bibinfo {year} {2019})},\ \Eprint
  {http://arxiv.org/abs/1907.11747} {arXiv:1907.11747 [hep-lat]} \BibitemShut
  {NoStop}%
\bibitem [{\citenamefont {Cali}\ \emph {et~al.}(2020)\citenamefont {Cali},
  \citenamefont {Cichy}, \citenamefont {Korcyl},\ and\ \citenamefont
  {Simeth}}]{Cali:2020hrj}%
  \BibitemOpen
  \bibfield  {author} {\bibinfo {author} {\bibfnamefont {S.}~\bibnamefont
  {Cali}}, \bibinfo {author} {\bibfnamefont {K.}~\bibnamefont {Cichy}},
  \bibinfo {author} {\bibfnamefont {P.}~\bibnamefont {Korcyl}}, \ and\ \bibinfo
  {author} {\bibfnamefont {J.}~\bibnamefont {Simeth}},\ }\href {\doibase
  10.1103/PhysRevLett.125.242002} {\bibfield  {journal} {\bibinfo  {journal}
  {Phys. Rev. Lett.}\ }\textbf {\bibinfo {volume} {125}},\ \bibinfo {pages}
  {242002} (\bibinfo {year} {2020})},\ \Eprint
  {http://arxiv.org/abs/2003.05781} {arXiv:2003.05781 [hep-lat]} \BibitemShut
  {NoStop}%
\bibitem [{\citenamefont {Bruno}\ \emph {et~al.}(2017)\citenamefont {Bruno},
  \citenamefont {Dalla~Brida}, \citenamefont {Fritzsch}, \citenamefont
  {Korzec}, \citenamefont {Ramos}, \citenamefont {Schaefer}, \citenamefont
  {Simma}, \citenamefont {Sint},\ and\ \citenamefont {Sommer}}]{Bruno:2017gxd}%
  \BibitemOpen
  \bibfield  {author} {\bibinfo {author} {\bibfnamefont {M.}~\bibnamefont
  {Bruno}}, \bibinfo {author} {\bibfnamefont {M.}~\bibnamefont {Dalla~Brida}},
  \bibinfo {author} {\bibfnamefont {P.}~\bibnamefont {Fritzsch}}, \bibinfo
  {author} {\bibfnamefont {T.}~\bibnamefont {Korzec}}, \bibinfo {author}
  {\bibfnamefont {A.}~\bibnamefont {Ramos}}, \bibinfo {author} {\bibfnamefont
  {S.}~\bibnamefont {Schaefer}}, \bibinfo {author} {\bibfnamefont
  {H.}~\bibnamefont {Simma}}, \bibinfo {author} {\bibfnamefont
  {S.}~\bibnamefont {Sint}}, \ and\ \bibinfo {author} {\bibfnamefont
  {R.}~\bibnamefont {Sommer}} (\bibinfo {collaboration} {ALPHA}),\ }\href
  {\doibase 10.1103/PhysRevLett.119.102001} {\bibfield  {journal} {\bibinfo
  {journal} {Phys. Rev. Lett.}\ }\textbf {\bibinfo {volume} {119}},\ \bibinfo
  {pages} {102001} (\bibinfo {year} {2017})},\ \Eprint
  {http://arxiv.org/abs/1706.03821} {arXiv:1706.03821 [hep-lat]} \BibitemShut
  {NoStop}%
\bibitem [{\citenamefont {Aoki}\ \emph {et~al.}(2009)\citenamefont {Aoki} \emph
  {et~al.}}]{PACS-CS:2009zxm}%
  \BibitemOpen
  \bibfield  {author} {\bibinfo {author} {\bibfnamefont {S.}~\bibnamefont
  {Aoki}} \emph {et~al.} (\bibinfo {collaboration} {PACS-CS}),\ }\href
  {\doibase 10.1088/1126-6708/2009/10/053} {\bibfield  {journal} {\bibinfo
  {journal} {JHEP}\ }\textbf {\bibinfo {volume} {10}},\ \bibinfo {pages} {053}
  (\bibinfo {year} {2009})},\ \Eprint {http://arxiv.org/abs/0906.3906}
  {arXiv:0906.3906 [hep-lat]} \BibitemShut {NoStop}%
\bibitem [{\citenamefont {Maltman}\ \emph {et~al.}(2008)\citenamefont
  {Maltman}, \citenamefont {Leinweber}, \citenamefont {Moran},\ and\
  \citenamefont {Sternbeck}}]{Maltman:2008bx}%
  \BibitemOpen
  \bibfield  {author} {\bibinfo {author} {\bibfnamefont {K.}~\bibnamefont
  {Maltman}}, \bibinfo {author} {\bibfnamefont {D.}~\bibnamefont {Leinweber}},
  \bibinfo {author} {\bibfnamefont {P.}~\bibnamefont {Moran}}, \ and\ \bibinfo
  {author} {\bibfnamefont {A.}~\bibnamefont {Sternbeck}},\ }\href {\doibase
  10.1103/PhysRevD.78.114504} {\bibfield  {journal} {\bibinfo  {journal} {Phys.
  Rev. D}\ }\textbf {\bibinfo {volume} {78}},\ \bibinfo {pages} {114504}
  (\bibinfo {year} {2008})},\ \Eprint {http://arxiv.org/abs/0807.2020}
  {arXiv:0807.2020 [hep-lat]} \BibitemShut {NoStop}%
\bibitem [{\citenamefont {Banerjee}\ \emph {et~al.}(2012)\citenamefont
  {Banerjee}, \citenamefont {Datta}, \citenamefont {Gavai},\ and\ \citenamefont
  {Majumdar}}]{Banerjee:2011ra}%
  \BibitemOpen
  \bibfield  {author} {\bibinfo {author} {\bibfnamefont {D.}~\bibnamefont
  {Banerjee}}, \bibinfo {author} {\bibfnamefont {S.}~\bibnamefont {Datta}},
  \bibinfo {author} {\bibfnamefont {R.}~\bibnamefont {Gavai}}, \ and\ \bibinfo
  {author} {\bibfnamefont {P.}~\bibnamefont {Majumdar}},\ }\href {\doibase
  10.1103/PhysRevD.85.014510} {\bibfield  {journal} {\bibinfo  {journal} {Phys.
  Rev. D}\ }\textbf {\bibinfo {volume} {85}},\ \bibinfo {pages} {014510}
  (\bibinfo {year} {2012})},\ \Eprint {http://arxiv.org/abs/1109.5738}
  {arXiv:1109.5738 [hep-lat]} \BibitemShut {NoStop}%
\bibitem [{\citenamefont {Karsch}\ \emph {et~al.}(2003)\citenamefont {Karsch},
  \citenamefont {Laermann}, \citenamefont {Petreczky},\ and\ \citenamefont
  {Stickan}}]{Karsch:2003wy}%
  \BibitemOpen
  \bibfield  {author} {\bibinfo {author} {\bibfnamefont {F.}~\bibnamefont
  {Karsch}}, \bibinfo {author} {\bibfnamefont {E.}~\bibnamefont {Laermann}},
  \bibinfo {author} {\bibfnamefont {P.}~\bibnamefont {Petreczky}}, \ and\
  \bibinfo {author} {\bibfnamefont {S.}~\bibnamefont {Stickan}},\ }\href
  {\doibase 10.1103/PhysRevD.68.014504} {\bibfield  {journal} {\bibinfo
  {journal} {Phys. Rev. D}\ }\textbf {\bibinfo {volume} {68}},\ \bibinfo
  {pages} {014504} (\bibinfo {year} {2003})},\ \Eprint
  {http://arxiv.org/abs/hep-lat/0303017} {arXiv:hep-lat/0303017} \BibitemShut
  {NoStop}%
\bibitem [{\citenamefont {Stickan}\ \emph {et~al.}(2004)\citenamefont
  {Stickan}, \citenamefont {Karsch}, \citenamefont {Laermann},\ and\
  \citenamefont {Petreczky}}]{Stickan:2003gh}%
  \BibitemOpen
  \bibfield  {author} {\bibinfo {author} {\bibfnamefont {S.}~\bibnamefont
  {Stickan}}, \bibinfo {author} {\bibfnamefont {F.}~\bibnamefont {Karsch}},
  \bibinfo {author} {\bibfnamefont {E.}~\bibnamefont {Laermann}}, \ and\
  \bibinfo {author} {\bibfnamefont {P.}~\bibnamefont {Petreczky}},\ }\href
  {\doibase 10.1016/S0920-5632(03)02654-9} {\bibfield  {journal} {\bibinfo
  {journal} {Nucl. Phys. B Proc. Suppl.}\ }\textbf {\bibinfo {volume} {129}},\
  \bibinfo {pages} {599} (\bibinfo {year} {2004})},\ \Eprint
  {http://arxiv.org/abs/hep-lat/0309191} {arXiv:hep-lat/0309191} \BibitemShut
  {NoStop}%
\end{thebibliography}
